\documentclass[journal ]{new-aiaa}
\usepackage[utf8]{inputenc}
\usepackage{textcomp}
\usepackage{xcolor}
\usepackage{savesym}
\savesymbol{iint}
\savesymbol{iiint}
\savesymbol{iiiint}
\savesymbol{idotsint}
\usepackage{amsfonts}
\usepackage{amsmath}
\usepackage{graphicx}
\usepackage{savesym}

\usepackage{graphicx}
\usepackage[version=4]{mhchem}
\usepackage{longtable,tabularx}
\usepackage[flushleft]{threeparttable}

\usepackage{cleveref}
\crefname{section}{§}{§§}
\Crefname{section}{§}{§§}

\makeatletter
\def\@fnsymbol#1{\ensuremath{\ifcase#1\or \dagger\or \dagger\dagger\or *\or
\mathsection\or \mathparagraph\or \|\or **\or \dagger\dagger
\or \ddagger\ddagger \else\@ctrerr\fi}}
\makeatother

\title{Compressible Velocity Transformations for Various Noncanonical Wall-Bounded Turbulent Flows}

\author{Tianyi Bai \footnote{Ph.D. Candidate, Department of Mechanical and Aerospace Engineering. The first two authors contributed equally.}}
\affil{The Hong Kong University of Science and Technology, Clear Water Bay, Kowloon, Hong Kong}

\author{Kevin P. Griffin\footnote{Ph.D. Candidate, Center for Turbulence Research. The first two authors contributed equally.}}
\affil{Stanford University, Stanford, CA, USA}

\author{Lin Fu\footnote{Assistant Professor, Department of Mechanical and Aerospace Engineering; Department of Mathematics; Shenzhen Research Institute (The Hong Kong University of Science and Technology, Shenzhen, China). Corresponding author. Email: linfu@ust.hk.}}
\affil{The Hong Kong University of Science and Technology, Clear Water Bay, Kowloon, Hong Kong \\}

\begin{document}

\maketitle

\begin{abstract}
While several velocity transformations for compressible zero-pressure-gradient (ZPG) boundary layers have been proposed in the past decades, their performance for non-canonical compressible wall-bounded turbulent flows has not been systematically investigated.
This work assesses several popular transformations for the velocity profile through their application to several types of non-canonical compressible wall-bounded turbulent flows. Specifically, this work explores DNS databases of high-enthalpy boundary layers with dissociation and vibrational excitation, supercritical channel and boundary-layer flows, and adiabatic boundary layers with pressure gradients. The transformations considered include the van Driest [Van Driest, J. Aeronaut. Sci., 18(1951):145-216], Zhang et al. [Zhang et al., Phys. Rev. Lett., 109(2012):054502], Trettel-Larsson [Trettel and Larsson, Phys. Fluids, 28(2016):026102], data-driven [Volpiani et al., Phys. Rev. Fluids, 5(2020):052602], and total-stress-based [Griffin et al., Proc. Natl. Acad. Sci. U.S.A., 118(2021):e2111144118] transformations. The Trettel-Larsson transformation collapses velocity profiles of high-enthalpy  {temporal} boundary layers  {but not the spatial boundary layers considered}. For supercritical channel flows, the Trettel-Larsson transformation also performs well over the entire inner layer. None of the transformations above works for supercritical boundary layers. For all the considered methods, the transformed velocity profiles of boundary layers with weak pressure gradients coincide well with the universal incompressible law of the wall. In summary, all these popular methods fail to deliver uniform performance for non-canonical compressible wall-bounded flows in the logarithmic region, and a more sophisticated version, which accounts for these different physics, is needed. The data-driven and total-stress-based transformations perform well in the viscous sublayer for all the considered flows. Nevertheless, the present assessment provides a useful guideline on the deployment of these transformations on various non-canonical flows.
\end{abstract}

\section*{Nomenclature}

\noindent\begin{tabular}{@{}lcl@{}}
\textit{$\rho$}  &=& density, $kg/m^3$ \\
\textit{$p$}  &=& pressure, $Pa$ \\
\textit{$T$} &=& temperature, $K$ \\
\textit{$u,v$}  &=& streamwise and wall-normal velocity, $m/s$ \\
\textit{$x$,$y$,$z$} &=& axes in the inertial coordinate system, $m$ \\
\textit{$u_\tau$} &=& friction velocity, $m/s$ \\
\textit{$\delta_v$} &=& viscous length scale, m \\
\textit{$\mu$} &=& dynamic viscosity, $Pa \cdot s$ \\
\textit{$\tau$} &=& shear stress, $N/m^2$ \\
\textit{${\rm Re}$} &=&  Reynolds number \\
\textit{${\kappa}$} &=& Kármán constant \\
\textit{$\beta_k$} &=& Rotta-Clauser parameter \\
\end{tabular} \\

%%%%%%%%%%%%%%%%%%%%%%%%%%%%%%%%%%%%%%%%%%%%%%%%%%%%%%%%%%%%%%
\section{Introduction}\label{into}
The incompressible zero-pressure-gradient wall-bounded turbulent flow is a canonical case that has been widely studied. In the inner layer, there exists an inner scaling for the velocity profile, which is called the law of the wall and was derived by Prandtl \cite{prandtl19257,prandtl1932turbulenten} and von Kármán \cite{von1930mechanische}, also known as the logarithmic law of the wall, which is valid for equilibrium, incompressible, constant-property flows at high Reynolds numbers. The logarithmic law has been a foundation in the wall-bounded turbulence field to validate simulations and experiments and to construct reduced-order near-wall models. As for compressible flows, the mean thermal properties become non-uniform and the direct deployment of the incompressible law of the wall fails to describe mean velocity profiles.

This motivates the idea of a compressible velocity transformation which maps a compressible velocity profile to the incompressible law of the wall \cite{van1951turbulent}. Similarly, a reliable compressible velocity transformation would be useful for the validation of simulations and experiments and would enable the development of new turbulence models in the near-wall region, which can provide accurate mean flow prediction with reduced computational cost, e.g., for the aerospace industry.

To extend the logarithmic law of incompressible flows to compressible flows, Morkovin \cite{Morkovin} first observes that at moderate Mach number, the effect of the density fluctuations is negligible and concludes that “the essential dynamics of these shear flows will follow the incompressible pattern”, which is known as Morkovin's hypothesis \cite{Morkovin}. This hypothesis implies that the effects of compressibility can be accounted for by the variation of mean flow properties. In subsequent studies, many transformations have been developed following Morkovin's hypothesis \cite{Morkovin}.

One of the most widely used transformations is the van Driest \cite{van1951turbulent} transformation, which is derived based on the assumption of a constant stress layer and Prandtl's mixing-length model. This transformation implies that the variation of mean density can account for the effect of compressibility, and it performs well for flows over adiabatic walls. 

Zhang et al. \cite{zhang2012mach} argue against the assumption that Prandtl's mixing length is Mach-number invariant. A new Mach-number-invariant scaling is obtained by considering the quasi-balance of the production and dissipation terms in the turbulent kinetic energy (TKE) equation. Combining the assumptions of constant total stress and a modified Prandtl’s mixing-length model, Zhang et al. \cite{zhang2012mach} derive a transformation based on turbulence equilibrium.

Huang et al. \cite{huang1995compressible} proposed the semi-local scaling and Coleman et al. \cite{coleman1995numerical} showed that when turbulence intensities are plotted versus the wall-normal coordinate in semi-local units, a collapse is observed across various Mach numbers. Trettel and Larsson \cite{trettel2016mean} make an improvement of the van Driest \cite{van1951turbulent} transformation and derive the Trettel-Larsson \cite{trettel2016mean} transformation based on the semi-local scaling and the constant stress layer assumption. Patel et al. \cite{patel2016influence} develop the same transformation based on the observation that the viscous stress in the near-wall region collapses better in the semi-local scaling. This transformation works in channel flows very well, indeed across the entire inner layer.

A data-driven transformation has been developed by Volpiani et al. \cite{volpiani2020data}, by postulating that the compressibility effect can be accounted for by the non-dimensionalization with undetermined power laws of mean density and mean viscosity. The power law coefficients are fitted using the DNS data of boundary layers. This transformation was shown to succeed for both adiabatic and diabatic ZPG boundary layers.

Griffin et al. \cite{griffin2021velocity} claim that there are two fundamental flow scalings in the inner layer, i.e., one for the viscous sublayer and the other for the logarithmic region. In the logarithmic region, Griffin et al. \cite{griffin2021velocity} generalize the quasi-equilibirum argument of Zhang et al. \cite{zhang2012mach} through introducing the semi-local scaling. In the viscous sublayer, the transformation is the same as the Trettel-Larsson transformation. Griffin et al. \cite{zhang2012mach} further combine these two scalings and propose the total-stress-based transformation, which works for all the considered channel, pipe, and ZPG boundary-layer flows regardless of the thermal boundary condition, Reynolds number and Mach number.

While these popular transformations have been validated comprehensively for canonical ZPG boundary-layer flows, their performance for non-canonical flows is rarely investigated. In this work, we assess them for a wide range of non-canonical flows, including high-enthalpy flows with dissociation and excitation, supercritical flows, and flows with pressure gradients. The paper is organized as follows. In \cref{meth}, several transformations and their derivations are introduced; the databases and their numerical configurations are outlined in \cref{data}; the numerical assessments are provided in \cref{vali} with detailed analyses; concluding remarks are given in \cref{conc}.
%%
%%%%%%%%%%%%%%%%%%%%%%%%%%%%%%%%%%%%%%%%%%%%%%%%%%%%%%%%%%%%%%
\section{Various transformation methods}\label{meth}
%%%%%%%%%%%%%%%%%%%%%%%%%%%%%%%%%%%%%%%%%%%%%%%%%%%%%%%%%%%%%%

In this section, several established transformation methods will be introduced comprehensively, as well as the embedded assumptions under which they are derived. The connections and differences between these transformations will be discussed in detail.

Hereafter, both the Reynolds ($\phi=\overline{\phi}+\phi'$) and Favre ($\phi=\widetilde{\phi}+\phi''$) decompositions are used, where $\overline{\phi}$ and $\widetilde{\phi}= \overline{\rho \phi}/\overline{\rho}$ denote the Reynolds- and Favre-averaged quantities, and $\phi'$ and $\phi''$ denote the fluctuations of the Reynolds and Favre decompositions, respectively.
%and $\phi'$ and $\phi''$ denote fluctuations about the Reynolds- and Favre-averaged quantities, respectively.

Throughout this paper, the superscript $^+$ indicates a non-dimensionalization by the friction velocity $u_\tau = \sqrt{\tau_w/\overline{\rho}_w}$, the viscous length scale $\delta_v=\overline{\mu}_w/(u_\tau \overline{\rho}_w)$, and $\rho_w$, where $\tau_w = \overline{\mu}_w (\partial \widetilde{u}/\partial y)|_w$, $\rho_w$, and $\mu_w$ are the shear stress, density, and dynamic viscosity evaluated at the wall, respectively. For instance, $\widetilde{u}^+=\widetilde{u}/u_\tau$, $y^+ = y/\delta_v$, $\overline{\mu}^+ = \overline{\mu}/\overline{\mu}_w$, $\overline{\rho}^+=\overline{\rho}/\overline{\rho}_w$, etc. The friction Reynolds number is defined as $Re_{\tau}=\rho_w u_\tau \delta / \mu_w$, where $\delta$ is the boundary layer thickness.

%%%%%%%%%%%%%%%%%%%%%%%%%%%%%%%
\subsection{The van Driest transformation \cite{van1951turbulent}}
The first successful transformation for compressible wall-bounded flows was proposed by van Driest \cite{van1951turbulent}. The assumptions that a constant stress layer exists, and that the viscous stress is negligible, imply
\begin{equation}
\widetilde{u''v''} \sim \left( \frac{\overline{\rho}_w}{\overline{\rho}} \right) u_{\tau}^2.
\label{VD1}
\end{equation}
Further assuming that Prandtl's mixing-length model for the Reynolds shear stress holds in a compressible flow, we have
\begin{equation}
\overline{\rho}_w u_{\tau}^2=\overline{\rho}l_m^2 \left(\frac{\partial \widetilde{u}}{\partial y} \right)^2,
\label{VD2}
\end{equation}
where $l_m$ denotes the characteristic length scale of eddies. Prandtl assumes that the size of eddies is considered to be proportional to their distance to the wall, leading to the relation $l_m=\kappa y$, where $\kappa$ is the von Kármán constant. Substituting %Eq.~(\ref{VD3})
the length scale of eddies above to Eq.~(\ref{VD2}), the van Driest \cite{van1951turbulent} transformation is obtained
\begin{equation}
u_{VD}^+=\frac{1}{\kappa}\log y^++C, 
\label{VD4}
\end{equation}
where
\begin{equation}
du_{VD}^+= \sqrt{ \frac{\overline{\rho}}{\overline{\rho}_w} } d\widetilde{u}^+,\quad
y^+=y\frac{u_{\tau}}{\overline{\nu}_w}.
\label{VD5}
\end{equation}
Here, $\overline{\nu}_w$ is the kinematic viscosity at the wall, and $u_{VD}^+$ is the velocity transformed by the van Driest transformation \cite{van1951turbulent}. 

The performance of the van Driest \cite{van1951turbulent} transformation has been assessed by experiments of turbulent boundary layers (Fernholz and Finley \cite{fernholz1977critical,Fernholz1980ACC}, Fernholz et al. \cite{fernholz1981further}), DNS of adiabatic ZPG boundary layers (Pirozzoli and Bernardini \cite{pirozzoli2013probing,pirozzoli2011turbulence}, Pirozzoli et al. \cite{pirozzoli2004direct}, Guarini et al. \cite{guarini2000direct}), isothermal ZPG boundary layers (Maeder et al. \cite{maeder2001direct}, Mart{\'i}n \cite{martin2007direct}), and isothermal (nearly adiabatic) ZPG boundary layers (Duan et al. \cite{duan2011P3direct}, Lagha et al. \cite{lagha2011numerical}). The success of van Driest \cite{van1951turbulent} transformation for adiabatic flows verifies the validity of Morkovin’s hypothesis \cite{Morkovin} and builds up the foundation for the development of more advanced transformations. On the other hand, it has also been widely recognized that the van Driest transformation does not function well for boundary-layer flows with heat transfer \cite{trettel2016mean,griffin2021velocity}.
%%%%%%%%%%%%%%%%%%%%%%%%%%%%%
\subsection{The Zhang et al. transformation \cite{zhang2012mach}}
Appealing to the established Mach-number-invariance of several quantities, Zhang et al. \cite{zhang2012mach} propose a velocity transformation for adiabatic boundary layers. Zhang et al. \cite{zhang2012mach} recognize that in the logarithmic region, the dissipation term reduces to $\overline{\mu}\overline{\omega'_i\omega'_i}$ at moderate Mach numbers, and are in the quasi-balance with the turbulence production term $-\overline{\rho}\widetilde{u''v''}(\partial\widetilde{u}/\partial y)$ in the TKE equation, and the ratio of these two terms is Mach-number-invariant. By further observing that the distributions of $\overline{\omega'_i\omega'_i}^+$ and $-\overline{\rho}^+\widetilde{u''v''}^+$ are independent on the Mach number variations, one can conclude that $(\partial \widetilde{u}/\partial y)^+/\overline{\mu}^+$ denoted by $S$ is Mach-number-invariant and obtain the Mach-number-invariant scaling of the mixing length
\begin{equation}
l_{Z}^+=(\sqrt{\overline{\rho}^+}\overline{\mu}^+)l_{m}^+,
\label{Z1}
\end{equation}
where $l_{m}^+=(-{\widetilde{u''v''}^{+})^{\frac{1}{2}}}/(\partial \widetilde{u}^+/\partial y^+)$.  By combining the Mach-number-invariant mixing length and the constant-stress-layer assumption, Zhang et al. \cite{zhang2012mach} derive the following velocity transformation
\begin{equation}
\begin{aligned}
&u_{Z}^+=\int_0^{\widetilde{u}^+}\frac{g}{\overline{\mu}^+}\, d\widetilde{u}^+ ,\\
&g=\frac{-\frac{S}{2}+\sqrt{(\frac{S}{2})^2+(1-\overline{\mu}^{+2}S)}}{(1-\overline{\mu}^{+2}S)},
\label{Z2}
\end{aligned}
\end{equation}
where $u_{Z}^+$ denotes the corresponding transformed velocity. Similar to the van Driest transformation, the approach of Zhang et al. only works well for wall-bounded turbulence with an adiabatic boundary condition \cite{griffin2021velocity}.

%%%%%%%%%%%%%%%%%%%%%%%%%%
\subsection{The Trettel-Larsson transformation \cite{trettel2016mean}}
In the log region of incompressible wall-bounded flows, the relation $du/dy=(\tau_w/\overline{\rho}_w)^{1/2}/(\kappa y)$ holds, leading to the well-known universal log law, i.e., $du^+/dy^+=1/(\kappa y^+$). For a compressible flow, the key variables in the logarithmic region of compressible flows are the wall shear $\tau_w$, the local flow density $\overline{\rho}$, and the wall-normal distance $y$ \cite{bradshaw1994turbulence}. By analogy to the incompressible flow, the mean shear stress in a compressible flow is  {assumed by Trettel and Larsson \cite{trettel2016mean} to follow}
\begin{equation}
\frac{d\widetilde{u}}{dy}=\frac{1}{\kappa y}(\frac{\tau_w}{\overline{\rho}})^{\frac{1}{2}},
\label{TL1}
\end{equation}
with the velocity scale $\sqrt{{\tau_w}/{\overline{\rho}}}$ according to dimension analysis.  {In order for the velocity to follow the incompressible law of the wall after the transformation is applied, Trettel and Larsson \cite{trettel2016mean} derive that the transformed velocity and length scale must follow} the condition
\begin{equation}
\frac{du_{TL}}{d\widetilde{u}}=\frac{y}{y_{TL}}(\frac{\overline{\rho}}{\overline{\rho}_w})^{\frac{1}{2}}\frac{dy_{TL}}{dy},
\label{TL2}
\end{equation}
where $u_{TL}$  {and $y_{TL}$} denote the ``incompressible'' velocity  {and wall-normal distance} mapped by the Trettel-Larsson \cite{trettel2016mean} transformation.
Then, Trettel and Larsson \cite{trettel2016mean} employ the constant-stress-layer assumption to find
\begin{equation}
\overline{\mu}_w\frac{du_{TL}}{dy_{TL}}=\overline{\mu}\frac{d\widetilde{u}}{dy}.
\label{TL3}
\end{equation}
The velocity gradient derived from the two conditions above should be equal to each other and yield the Trettel-Larsson transformation \cite{trettel2016mean}

\begin{equation}
\begin{aligned}
&u_{TL}^+=\int_0^{\widetilde{u}^+} (\frac{\overline{\rho}}{\overline{\rho}_w})^{\frac{1}{2}}[1+\frac{1}{2}\frac{1}{\overline{\rho}}\frac{d\overline{\rho}}{dy}y-\frac{1}{\overline{\mu}}\frac{d\overline{\mu}}{dy}y]\, d\widetilde{u}^+ , \\
&y^*=\frac{\overline{\rho}(\tau_w/\overline{\rho})^{1/2}y}{\overline{\mu}},
\label{eq:TL4}
\end{aligned}
\end{equation}
where $y^*$ is often called the semi-local scaling of the wall-normal coordinate.

Trettel and Larsson \cite{trettel2016mean} validate this transformation with a compilation of experimental data of turbulent boundary layers from Fernholz and Finley \cite{fernholz1977critical} and Fernholz et al. \cite{fernholz1981further}, the DNS data of channel flows by Trettel and Larsson \cite{trettel2016mean}, and  {one diabatic low}-enthalpy  {temporal} boundary layer from Duan and Mart{\'i}n \cite{duan2011p4direct}. This transformation satisfactorily collapses the velocity profile in the entire inner layer of channel flows. However, recent research reveals that it typically fails in the log region when deployed to ZPG boundary layer flows with heat transfer at the wall \cite{griffin2021velocity}\cite{zhang2018direct}.

%%%%%%%%%%%%%%%%%%%%%%%%%%%%%
\subsection{The data-driven transformation \cite{volpiani2020data}}
Following Morkovin \cite{Morkovin}, Volpiani et al. \cite{volpiani2020data} postulate that the compressibility effect can be accounted for by the variation of mean density and mean viscosity, and they further assume the transformation is in the integral form
\begin{equation}
y_V=\int_0^{y}f_I\,dy,\quad
u_V=\int_0^{\widetilde{u}}g_I\,d\widetilde{u},
\label{V1}
\end{equation}
where $f_I$ and $g_I$ are mapping functions, $y_V$ and $u_V$ denote the resulting distributions from transformation $I$, and these profiles should agree with incompressible distributions if the transformation is successful. To be in line with the simplified constant-stress-layer assumption \cite{trettel2016mean} in the viscous sublayer, $f_I$ and $g_I$ must satisfy the relation $(\overline{\mu}/{\overline{\mu}_w})f_I=g_I$, which implies
\begin{equation}
f_I=(\frac{\overline{\rho}}{\overline{\rho}_w})^b(\frac{\overline{\mu}}{\overline{\mu}_w})^{-a},\quad
g_I=(\frac{\overline{\rho}}{\overline{\rho}_w})^b(\frac{\overline{\mu}}{\overline{\mu}_w})^{1-a}.
\label{V2}
\end{equation}

Following the data-driven philosophy, the exponent parameters $a$ and $b$ are calibrated by the DNS databases of adiabatic and diabatic boundary layers from Volpiani et al. \cite{volpiani2018effects,volpiani2020effects}, diabatic boundary layers from Volpiani et al. \cite{volpiani2020data}, adiabatic and diabatic boundary layers from Zhang et al. \cite{zhang2018direct}, and a channel flow from Modesti and Pirozzoli \cite{modesti2016reynolds}, and the optimal values turn out to be $3/2$ and $1/2$, respectively. Therefore, the data-driven velocity transformation is given as
\begin{equation}
\begin{aligned}
&u_V^+=\int_0^{\widetilde{u}^+}(\frac{\overline{\rho}}{\overline{\rho}_w})^\frac{1}{2}(\frac{\overline{\mu}}{\overline{\mu}_w})^{-\frac{1}{2}}\,d\widetilde{u}^+,\\ &y_V^+=\int_0^{y^+}(\frac{\overline{\rho}}{\overline{\rho}_w})^\frac{1}{2}(\frac{\overline{\mu}}{\overline{\mu}_w})^{-\frac{3}{2}}\,dy^+,
\label{V2}
\end{aligned}
\end{equation}
where $u_V^+$ and $y_V^+$ denote the transformed velocity and wall-normal distance function. This transformation works well for boundary layers, but its performance deteriorates for channel flows \cite{griffin2021velocity}. 
%%%%%%%%%%%%%%%%%%%%%%%%%%%%%
\subsection{The total-stress-based transformation \cite{griffin2021velocity}}
Inspired by the work mentioned above, Griffin et al. \cite{griffin2021velocity} argue that there should be different scalings in the viscous sublayer and the logarithmic region, respectively. In the logarithmic region, Griffin et al. \cite{griffin2021velocity} first generalize the quasi-equilibirum argument by Zhang et al. \cite{zhang2012mach} through introducing the semi-local scaling. Griffin et al. \cite{griffin2021velocity} recognize the ratio of turbulence production and viscous dissipation as a Mach-number-invariant function of $y^*$. This weaker assumption leads to the following dimensionless mean shear, i.e.,
\begin{equation}
S_{eq}^+(y^*)=\frac{1}{\overline{\mu}^+}\frac{\partial {\widetilde{u}^+}}{\partial {y^*}}=\frac{\partial {u_{eq}^+}}{\partial {y^*}},
\label{eq:T1}
\end{equation}
which is Mach-number-independent in the log region. 

Within the viscous sublayer, the viscous stress is assumed to be Mach-number-independent, leading to the non-dimensionalized mean shear
\begin{equation}
S_{TL}^+(y^*)=\overline{\mu}^+\frac{\partial {\widetilde{u}^+}}{\partial {y^+}}=\frac{\partial {u_{TL}^+}}{\partial {y^*}},
\label{eq:T2}
\end{equation}
which is also known as the Trettel-Larsson transformation \cite{trettel2016mean}.

To combine these two scalings to derive a composite transformation,  Griffin et al. \cite{griffin2021velocity} rewrite the total stress $\tau^+=\tau_v^++\tau_R^+$ as
\begin{equation}
\tau^+=S_t^+(\frac{\tau_v^+}{S_{TL}^+}+\frac{\tau_R^+}{S_{eq}^+}),
\label{T3}
\end{equation}
where $\tau^+$ denotes the non-dimensional total stress, $\tau_v^+$ the viscous stress, $\tau_R^+$ the Reynolds shear stress, and $S_t^+$ is a generalized non-dimensional mean shear, which is defined such that the equation holds. 
This construction allows $S_t^+$ to recover $S_{TL}^+$ as $\tau^+ \rightarrow \tau_v^+$ in the viscous sublayer and to recover $S_{eq}^+$ as $\tau^+ \rightarrow \tau_R^+$ in the logarithmic region. Noticing that $\tau_v^+=S_{TL}^+$, and plugging $\tau_R^+=\tau^+-\tau_v^+$ into Eq.~(\ref{T3}), $S_t^+$ can be analytically written as
\begin{equation}
S_t^+=\frac{\tau^+S_{eq}^+}{\tau^++S_{eq}^+-S_{TL}^+},
\label{eq:T4}
\end{equation}
and the transformed velocity $u_t^+$ is defined correspondingly as
\begin{equation}
u_t^+=\int_0^{y^*} S_t^+\, dy^*.
\label{T5}
\end{equation}
Based on the constant-stress-layer assumption, a simplified version can be derived with a similar performance, see Eq.~(5) of \cite{griffin2021velocity}.

This total-stress-based transformation \cite{griffin2021velocity} has been assessed by DNS databases of channel flows (Trettel and Larsson \cite{trettel2016mean}, Modesti and Pirozzoli \cite{modesti2016reynolds}, Yao and  Hussain \cite{yao2020turbulence}), pipe flows (Modesti and Pirozzoli \cite{modesti2019direct}), both adiabatic boundary layers (Pirozzoli and Bernardini \cite{pirozzoli2011turbulence},  Zhang et al. \cite{zhang2018direct}, Volpiani et al. \cite{volpiani2018effects}) and diabatic boundary layers (Zhang et al. \cite{zhang2018direct}, Volpiani et al. \cite{volpiani2018effects,volpiani2020effects}), and the high-Mach turbulent boundary layers downstream of the shock/boundary-layer interaction (Fu et al. \cite{fu2021shock}). These extensive numerical validations reveal that this new transformation performs very well for all the considered canonical wall-bounded flows, irrespective of the Mach number, Reynolds number, and the thermal boundary condition.
%%%%%%%%%%%%%%%%%%%%%%%%%%%%%%%%%%%%%%%%%%%%%%%%%%%%%%%%%%%%%%
\section{Database descriptions}\label{data}
The DNS databases utilized to assess different velocity transformations include high-enthalpy  {and low-enthalpy} turbulent  {temporal} boundary layers with dissociation and vibrational excitation by Duan and Mart{\'i}n \cite{duan2011p4direct},  {the high-enthalpy spatial boundary layer of di Renzo and Urzay \cite{renzo2021direct}}, diabatic channel flows at supercritical pressure by Wan et al. \cite{wan2020mean} and Toki et al. \cite{toki2020velocity}, heated and unheated boundary layers at supercritical pressure by Kawai \cite{kawai2019heated}, and adiabatic boundary layers with pressure gradients by Wenzel et al. \cite{wenzel2019self} and Gibis et al. \cite{gibis2019self}.
\subsection{High-enthalpy turbulent boundary layers}
For flying vehicles at a high Mach number, the shock  {heating} can lead to extremely hot boundary layers downstream. This high-enthalpy condition can cause internal excitation and dissociation of air molecules. The  {specific} heat capacity varies  {with temperature} as a consequence, and both the thermal and chemical equilibrium  {may be} disrupted.

Duan and Mart{\'i}n \cite{duan2011p4direct} conduct the DNS of high- and low-enthalpy wall-bounded flows.  {Zero-pressure-gradient temporal boundary-layer simulations are conducted for various enthalpy conditions and Mach numbers. The two low enthalpy cases assume a calorically perfect gas. The four high-enthalpy cases match the conditions downstream of the oblique shock for wedges of} $35^{\circ}$ and $8^{\circ}$ (denoted by Wedge35 and Wedge8). To investigate the effect of species boundary conditions, ‘supercatalytic’ and ‘non-catalytic’ surface models are considered  {for each wedge angle}. For the non-catalytic model, there is no atom recombination and there exists minimal enthalpy recovery at the wall. It can be written as $(\partial Y/\partial n)_w=0$, where $Y$ is the species concentration and $n$ is the wall-normal coordinate. On the contrary, the supercatalytic model means infinitely fast atom recombination and maximum enthalpy recovery implying $Y_w=Y_{\infty}$. For the low-enthalpy flows, their edge Mach number and the ratio of wall to recovery temperature are carefully chosen to approximately match those for high-enthalpy cases. The corresponding low-enthalpy cases are denoted as LowH\_M3 and LowH\_M10 for Wedge35 and Wedge8, respectively.

The wall  {boundary conditions are} no-slip and isothermal  {(cold wall)}. The flow variables on the top boundary are fixed and obtained from a RANS calculation  {of a larger domain that includes the wedge and associated oblique shock}. Periodic conditions are enforced for the streamwise and spanwise boundaries.

 {The DNS of di Renzo and Urzay \cite{renzo2021direct} is a spatially evolving boundary layer which includes the transition from a laminar to a fully turbulent boundary layer. Only profiles from the fully turbulent section of the domain ($Re_\tau > 600$) are considered in this work. A non-catalytic, flat, cold wall is used. The freestream Mach number is 10, and the stagnation enthalpy is 21.6 MJ/kg, which corresponds to the post-shock conditions of a $9^\circ$ wedge flying at Mach 23 in Earth's atmosphere at 25 km. Vibrational excitation and dissociation are considered.}

%%%%%%%%%%%%%%%%%%
\subsection{Flows at supercritical pressure}
%%%%%%%%%%%%%%%%%%
\subsubsection{Channel flows at supercritical pressure}
Toki et al. \cite{toki2020velocity} conduct DNSs of planar turbulent channel flows driven by a body force with a wall-normal temperature gradient. The working fluid is nitrogen at a supercritical pressure, which is approximately 1.3 times its critical pressure. Since the ideal-gas assumption becomes invalid, the Soave-Redlich-Kwong (SRK) \cite{soave1972equilibrium} equation of state is employed. The bottom wall (116 K) is colder than its pseudocritical temperature (131.8 K at 4.4 MPa) while the top wall is hotter. Toki et al. \cite{toki2020velocity} evaluate the thermal conductivity $\lambda$ and the viscosity coefficient $\mu$ using the formula by Chung \cite{poling2001properties}. Two cases named R4 and R8 are considered, whose density ratio of the two walls is 4 and 8, respectively, and their corresponding temperatures at the top wall are 149 K and 219 K. The friction Reynolds number $Re_{\tau}$ is approximately 390 at the cold wall, 268 and 122 at the top wall for R4 and R8 respectively. The channel is $3\pi H$ in length and  $\pi H$ in the spanwise direction, where $H$ is the half-height of the channel. The no-slip condition is applied on the top and bottom walls. Periodic boundary conditions are applied in both the spanwise and streamwise directions.

Wan et al. \cite{wan2020mean} perform a similar investigation. The working fluid is carbon dioxide at 8 MPa and the corresponding pseudocritical temperature $T_{pc}=307.8$ K. The thermodynamic properties are taken from the property table of the REFPROF 9.1 database \cite{lemmon2002nist}. The channel is $4 H$ and $16 H$ in the spanwise and streamwise directions, respectively, where $H$ is the half-height of the channel. The no-slip boundary condition is applied at both walls with constant temperature ($T_{pc}-T_{w, cold}=T_{w, hot}-T_{pc}$). Periodic boundary conditions are applied in both the spanwise and streamwise directions. Wan et al. \cite{wan2020mean} conduct 5 simulations whose temperature difference between the two walls increases and one extra case with constant properties as a reference (only three of them will be considered in this work). The maximum temperature difference is selected to cover the significant variation range of all properties. All cases share the same bulk Reynolds numbers $Re_b=5705$.
%%%%%%%%%%%%%
\subsubsection{Boundary layers at supercritical pressure}
Kawai \cite{kawai2019heated} conducts boundary layer simulations at two different supercritical pressures, i.e., 2 MPa and 4 MPa, for the substance parahydrogen, and their corresponding pseudocritical temperatures are approximately 36.4 K and 43.3 K, respectively. Thermodynamic and fluid transport properties are evaluated using REFPROP (NIST Standard Reference Database 23, Version 9.0). The flat plate consists of two parts, i.e., the unheated part ($0 < x < 20 \delta_0$) and the heated part ($20 \delta_0 < x < 45 \delta_0$), where $\delta_0$ is approximately
99$\%$ of the boundary layer thickness at the inlet of the heated domain. Six boundary layer profiles are considered, i.e., two of them from unheated section of the wall ($T_w = T_\infty=25$ K) and the rest four from the heated section ($T_w = 100$ K and 200 K). The plate is isothermal and no-slip, and the periodic condition is adopted  for the spanwise boundaries. The friction Reynolds numbers are $Re_{\tau}\approx390$ for the unheated cases, $Re_{\tau}\approx$ 210 and 90 for cases with $T_w= 100$ K and 200 K at the pressure of 2 Mpa, and $Re_{\tau}\approx$ 360 and 150 for cases with $T_w= 100$ K and 200 K at the pressure of 4 Mpa, respectively.
%%%%%%%%%%%%%%%%%%%%%%%%%%%%%
\subsection{Boundary layers with pressure gradients}
The DNS of boundary layers with adverse (APG) and favourable (FPG) pressure gradients have been systematically explored by Wenzel et al. \cite{wenzel2019self} and Gibis et al. \cite{gibis2019self}. The kinematic Rotta-Clauser parameter $\beta_k=(\delta^*/\tau_w)(d\overline{p}_e/dx)$, where $\delta_K^*=\int_0^{\delta_{99}}(1-\overline{u}/\overline{u}_e)\, dy$ denotes the kinematic displacement thickness, and $d\overline{p}_e/dx$ the pressure gradient at the edge of the boundary layer, is adopted as the appropriate parameter representing the streamwise self-similarity for boundary layers with pressure gradients. Both subsonic ($M_{\infty,0}=0.5$) and supersonic ($M_{\infty,0}=2$) flows are considered, where $M_{\infty,0}$ is the free-stream Mach number at the inlet. For subsonic APG boundary layers, Rotta-Clauser parameters are set to be $\beta_k=$0, 0.19, 0.58, and 1.05, respectively. For hypersonic cases, Rotta-Clauser parameters are $\beta_k=$0, 0.15, 0.42, 0.55 and 0.62 for APG boundary layers and -0.18 for FPG boundary layer. For each simulation, the Rotta-Clauser parameter remains constant in the streamwise direction. The adiabatic no-slip condition is applied to the wall, and the periodic condition is adopted for spanwise boundaries.
%%%%%%%%%%%%%%%%%%%%%%%%%%%%%%
%%%%%%%%%%%%%%%%%%%%%%%%%%%%%%%%%%%%%%%%%%%%%%%%%%%%%%%%%%%%%%
\section{Assessment and analysis}\label{vali}
In this section, various transformations will be assessed by the databases mentioned above. Also, distributions of the various non-dimensionalizations of the mean shear will be analyzed as diagnostic functions indicating the accuracy of the transformations as a function of wall-normal distance. The results are presented case by case and then summarized in Tables \ref{tab:1} and \ref{tab:2} afterwards.

%%%%%%%%%%%%%%%%%%%%%%%%%%%%%%
\subsection{High-enthalpy turbulent boundary layers}
\begin{figure*}
    \centering
    \includegraphics[width=0.9\linewidth]{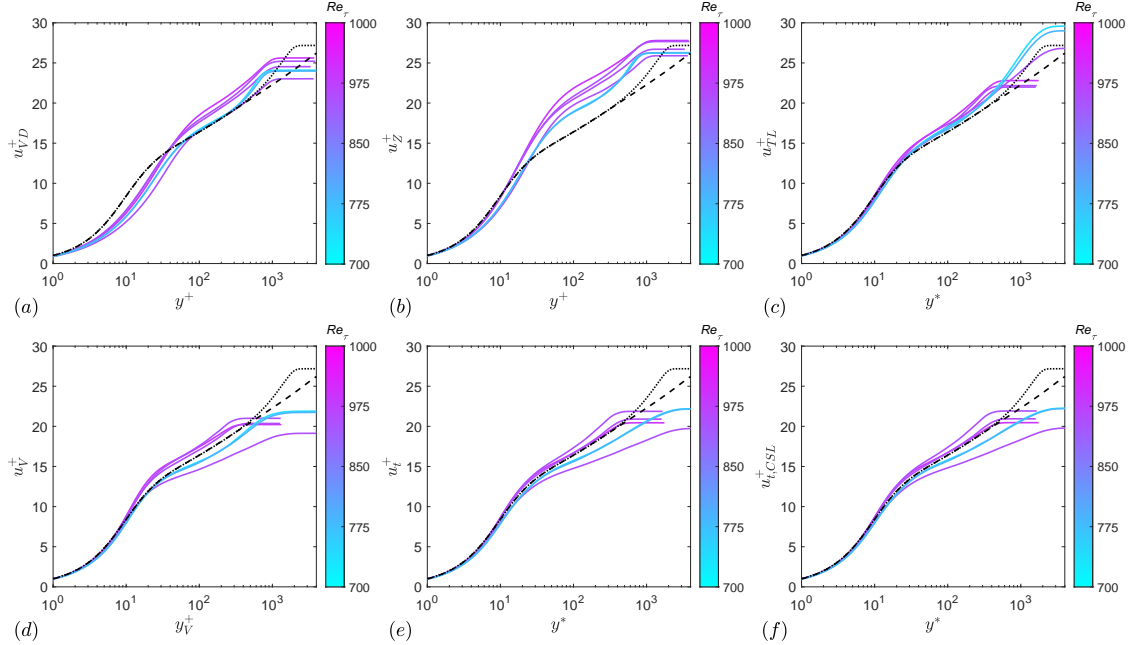}
    \caption{ {Transformed velocity distributions versus the non-dimensional wall-normal coordinate. The database of high-enthalpy  {and low-enthalpy} turbulent  {temporal} boundary layers by Duan and Mart{\'i}n \cite{duan2011p4direct} is used.}}
    %\caption{Transformed velocity distributions versus the non-dimensional wall-normal coordinate with the van Driest \cite{van1951turbulent} (a), Zhang et al. \cite{zhang2012mach} (b), Trettel-Larsson \cite{trettel2016mean} (c), data-driven \cite{volpiani2020data} (d), total-stress-based \cite{griffin2021velocity} (e), and constant-stress-layer version of the total-stress-based \cite{griffin2021velocity} (f) transformations. The database of high-enthalpy turbulent boundary layers by Duan and Mart{\'i}n \cite{duan2011p4direct} is used. The color of lines indicates the friction Reynolds number $Re_{\tau}$. Two incompressible DNS mean velocity profiles, i.e., from the channel flow (black dashed line) by Lee and Moser \cite{lee2015direct} at $Re_{\tau}\approx5200$ and the ZPG boundary layer (black dotted line) by Sillero et al. \cite{sillero2013one} at $Re_{\tau}\approx2000$, are also shown for comparisons.} 
    \label{fig:vel_duan}
\end{figure*}
%%
%%%
\begin{figure*}
    \centering
    \includegraphics[width=0.32\linewidth]{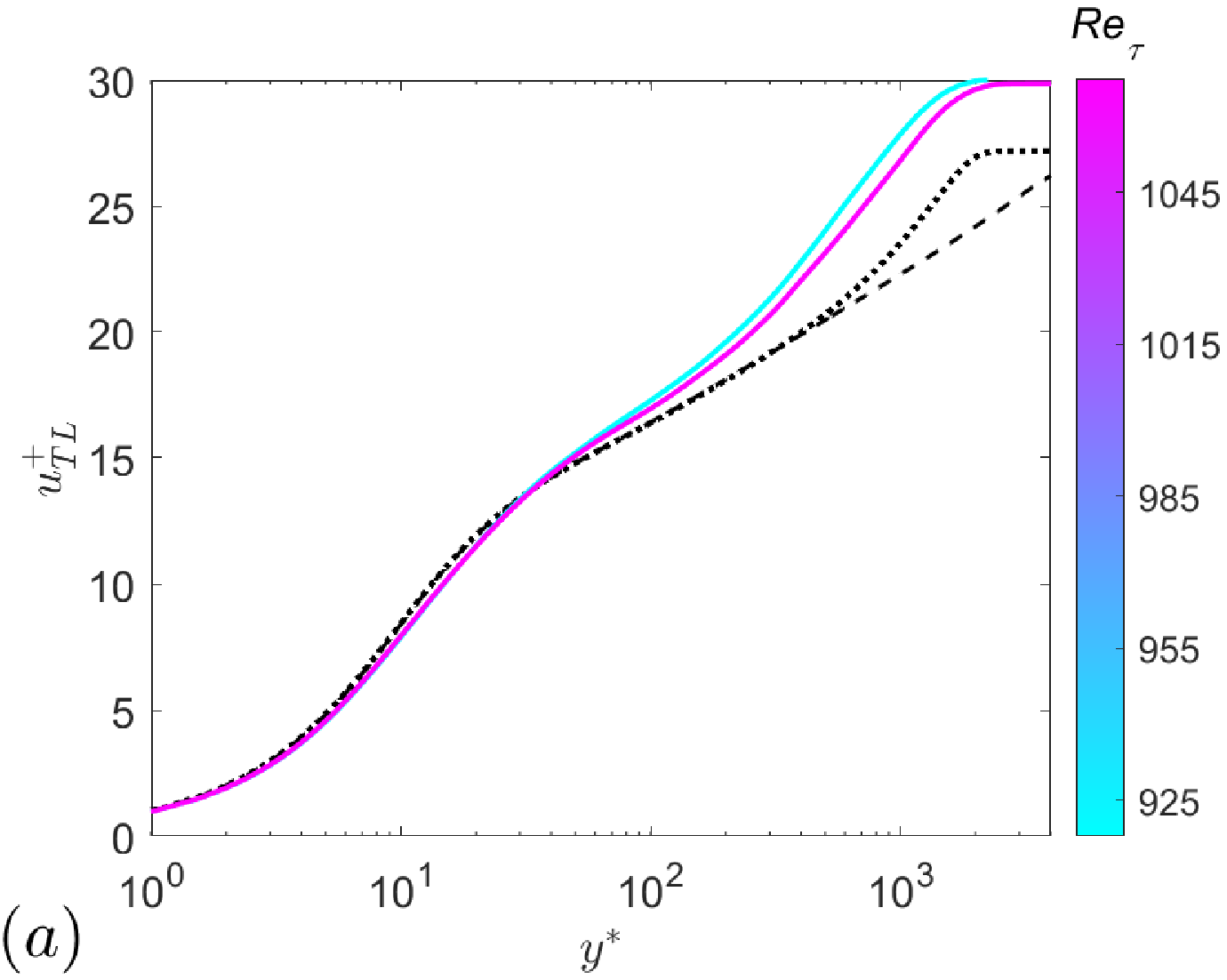}
    \includegraphics[width=0.32\linewidth]{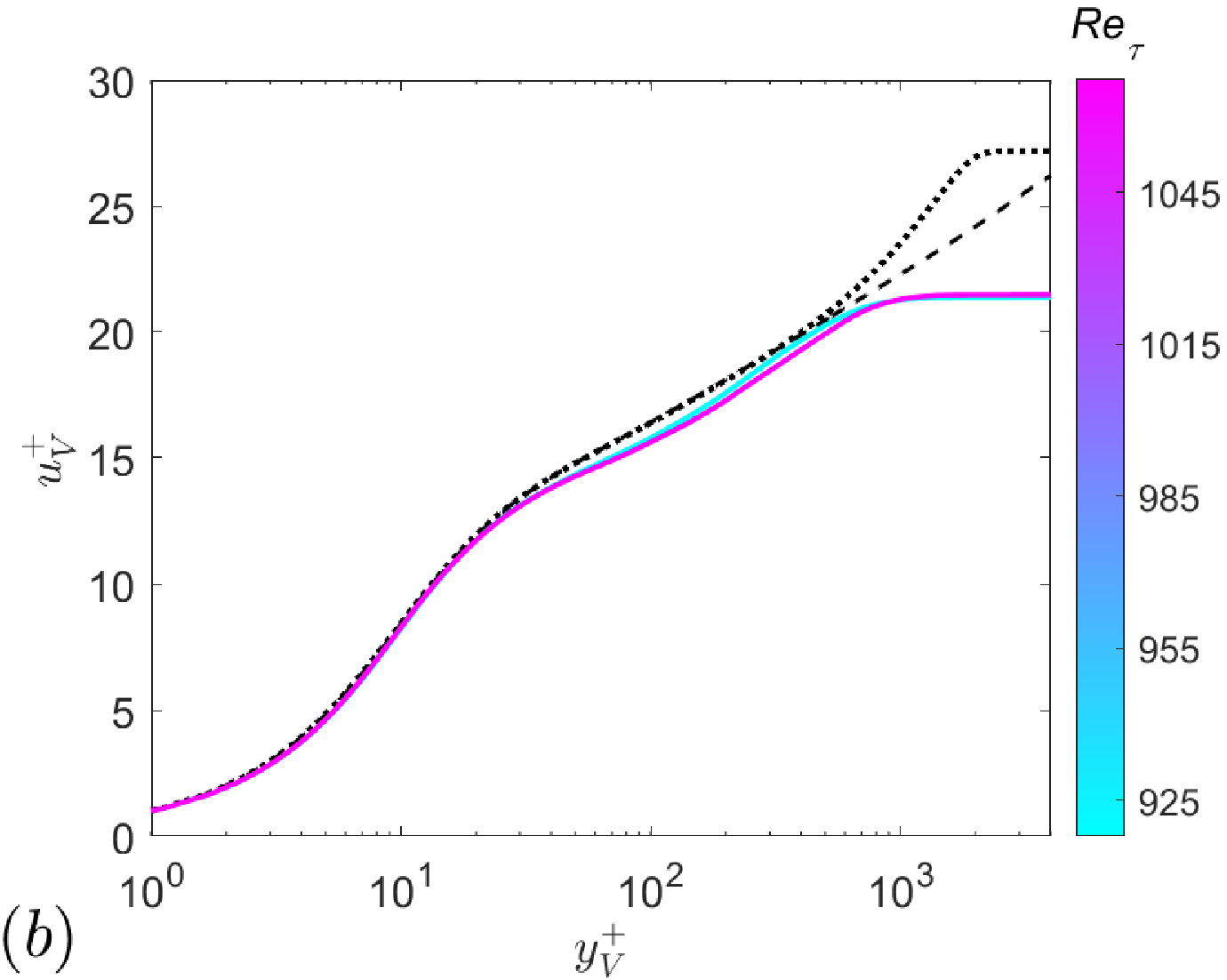}
    \includegraphics[width=0.32\linewidth]{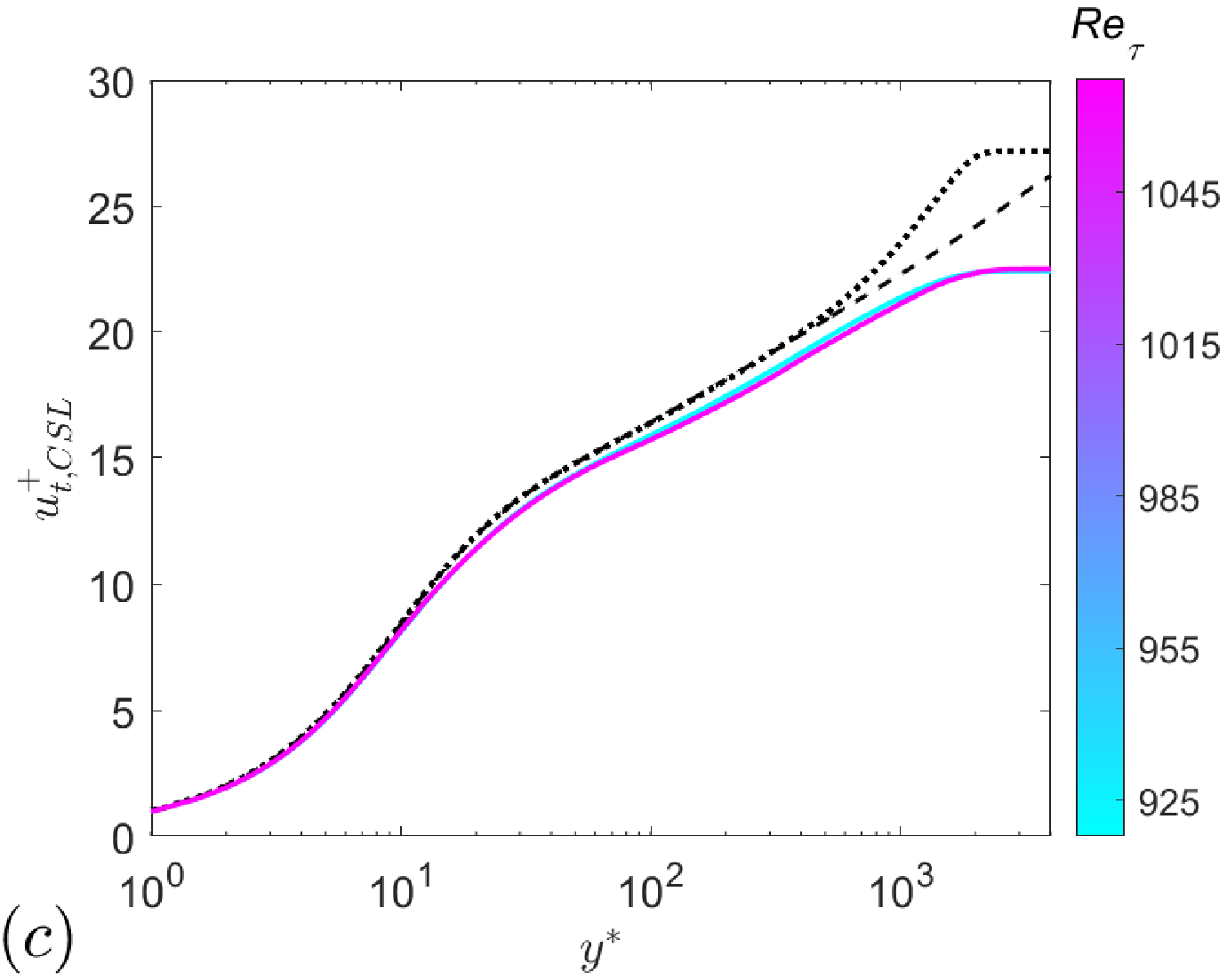}
    \caption{ {Transformed velocity distributions versus the non-dimensional wall-normal coordinate. The database of high-enthalpy turbulent spatial boundary layers by di Renzo and Urzay \cite{renzo2021direct} is used.}}
    \label{fig:vel_renzo}
\end{figure*}
%%%
%%
\begin{figure*}
    \centering
    \includegraphics[width=0.45\linewidth]{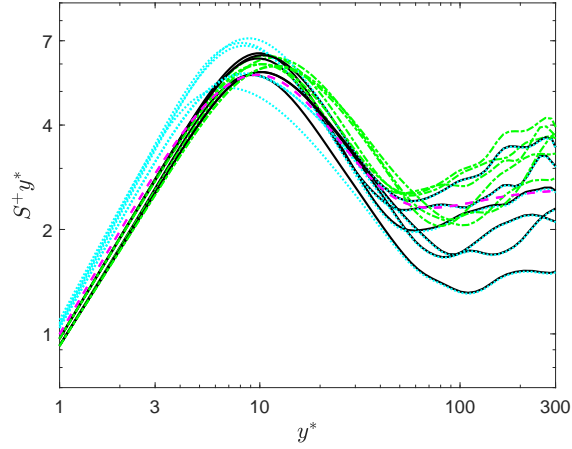}
    \caption{ {Three types of non-dimensional mean shear multiplied by the semi-local wall-normal coordinate are plotted with respect to the semi-local wall-normal coordinate. The database of high-enthalpy  {and low-enthalpy} turbulent  {temporal} boundary layers by Duan and Mart{\'i}n \cite{duan2011p4direct} is used.}}
    %\caption{Three types of non-dimensional mean shear multiplied by the semi-local wall-normal coordinate are plotted with respect to the semi-local wall-normal coordinate, where the one derived from Trettel-Larsson \cite{trettel2016mean} is in green, the one based on turbulence quasi-equilibrium in cyan (see Eq.~(\ref{eq:T1})), the one from the total-stress-based \cite{griffin2021velocity} transformation in black (see Eq.~(\ref{eq:T4})), and the last one $y^+(\partial u^+/\partial y^+)$ for the incompressible channel flow by Lee and Moser \cite{lee2015direct} at $Re_{\tau}\approx5200$ in magenta. The database of high-enthalpy turbulent boundary layers by Duan and Mart{\'i}n \cite{duan2011p4direct} is used.}
    \label{fig:diag_duan}
\end{figure*}
%
%%%
\begin{figure*}
    \centering
    \includegraphics[width=0.45\linewidth]{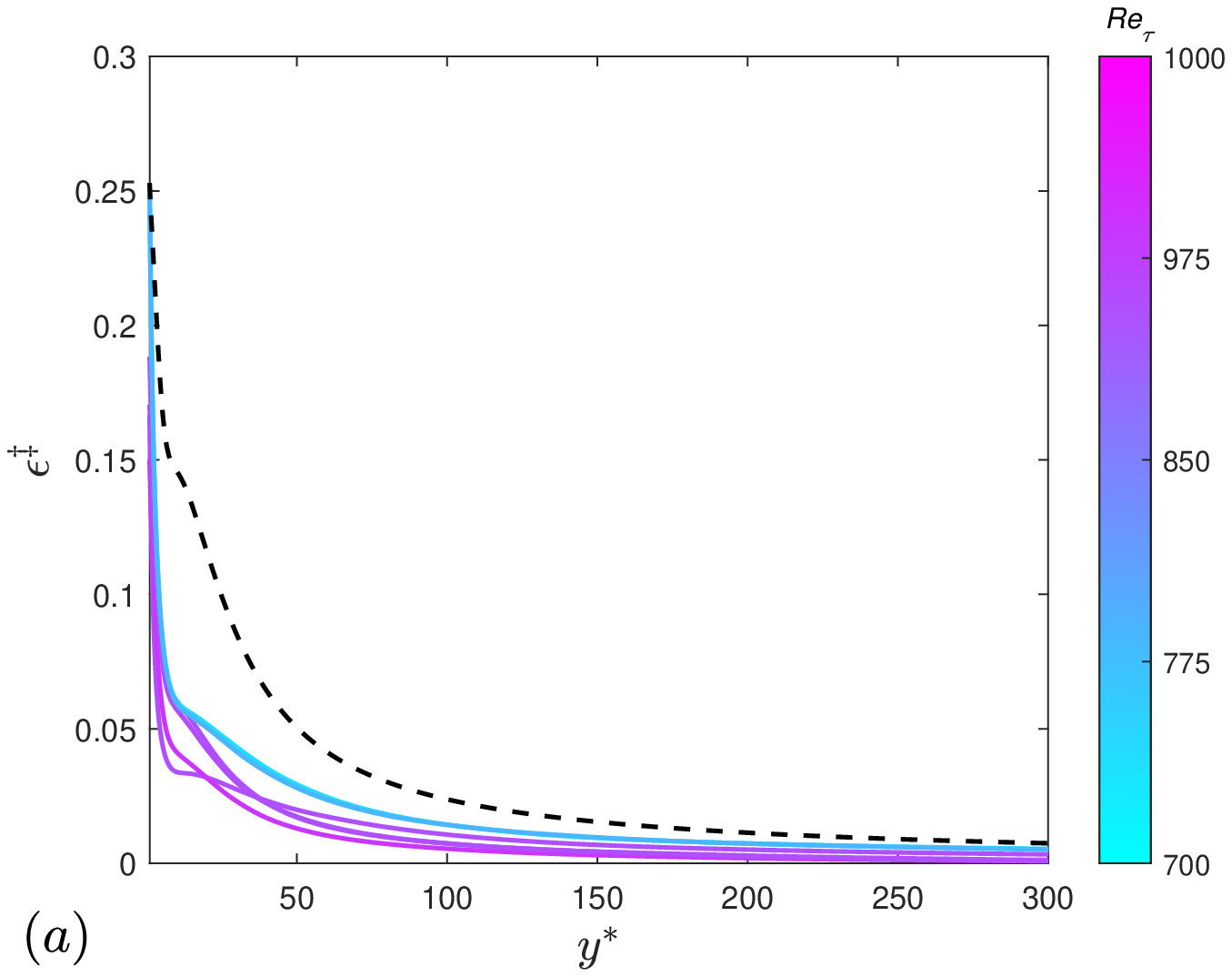}
    \includegraphics[width=0.45\linewidth]{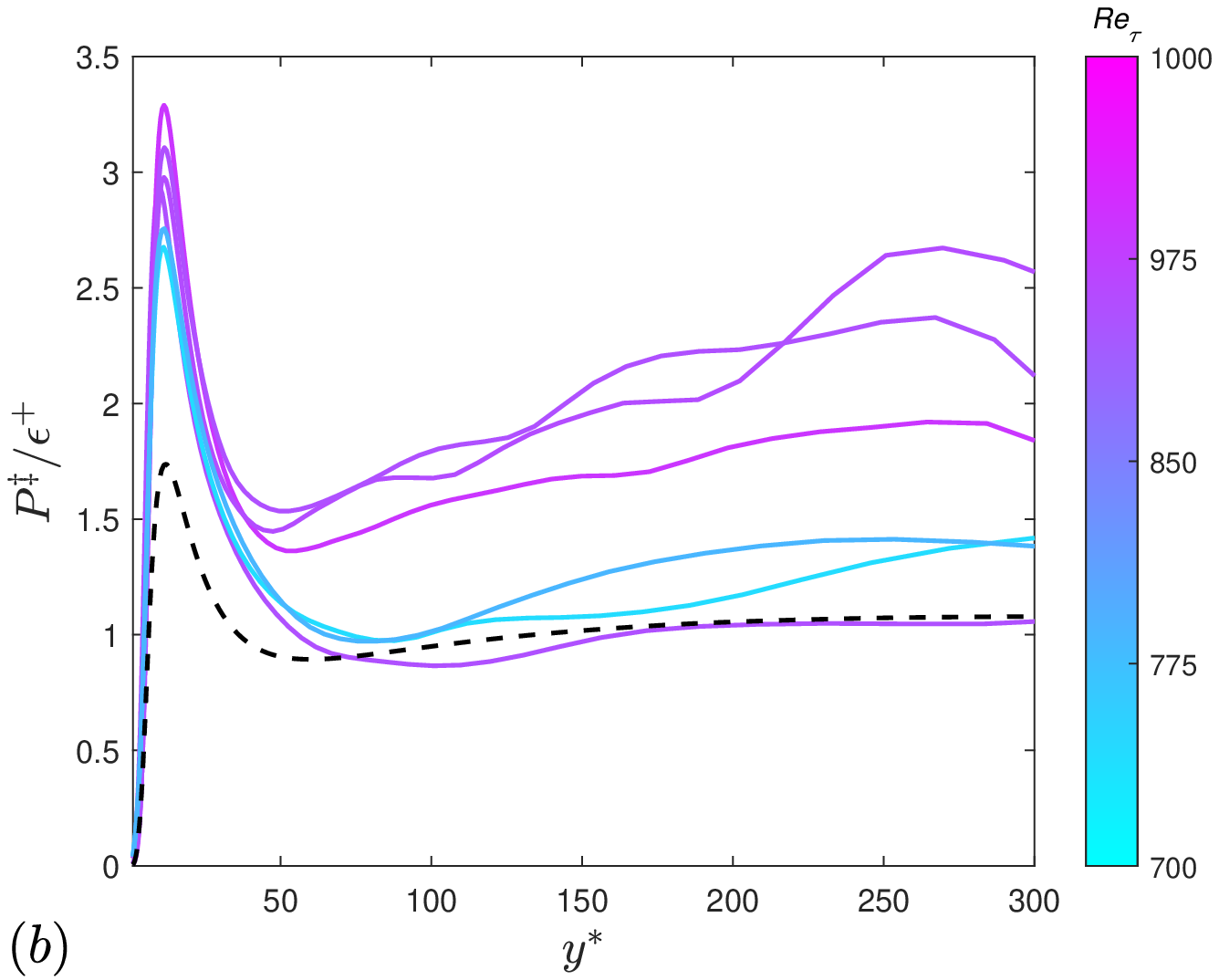}
    \caption{The quantity $\epsilon^{\ddagger}=\epsilon^+/\mu^+$ (a) and the ratio of the modified production $P^{\ddagger}=-\overline{\rho}^+\overline{u'v'}^+(\partial\widetilde{u}^+/\partial y^*)$ and dissipation \cite{griffin2021velocity} (b). The incompressible channel flow by Lee and Moser \cite{lee2015direct} at $Re_{\tau}\approx5200$ is employed as reference in the dashed black line. The database of high-enthalpy  {and low-enthalpy} turbulent  {temporal} boundary layers by Duan and Mart{\'i}n \cite{duan2011p4direct} is used.}
    \label{fig:diss_Duan}
\end{figure*}
%%%
%%%
\begin{figure*}
    \centering
    \includegraphics[width=0.45\linewidth]{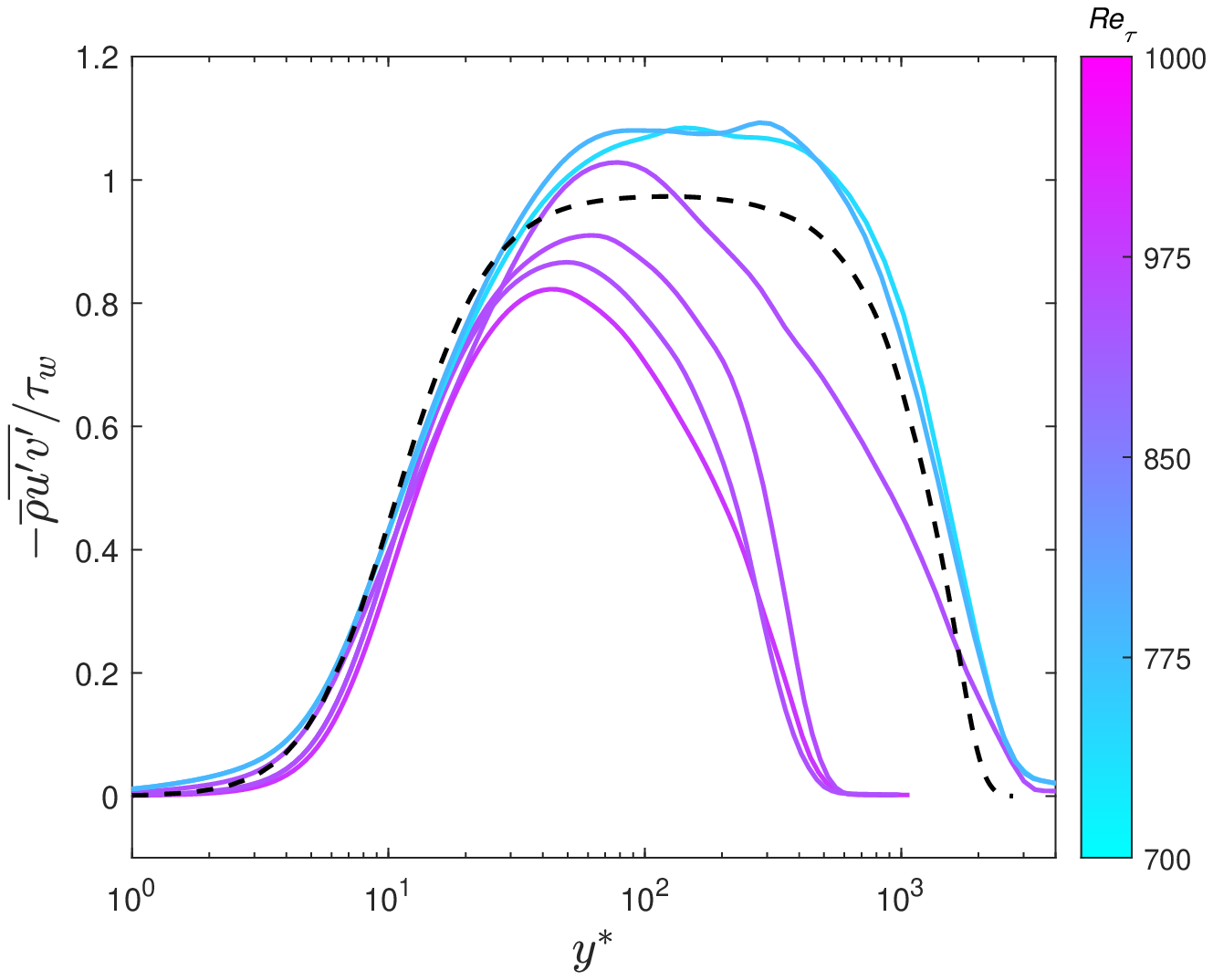}
    \caption{The approximate Reynolds shear stress profiles normalized by the wall stress for the high-enthalpy  {and low-enthalpy} turbulent boundary layers by Duan and Mart{\'i}n \cite{duan2011p4direct}. The incompressible boundary layer by Sillero et al. \cite{sillero2013one} at $Re_{\tau}\approx2000$ is employed as reference in the dashed black line.}
    \label{fig:RS_Duan}
\end{figure*}
%%%
 {In Fig.~\ref{fig:vel_duan}, transformed velocity distributions are plotted versus the non-dimensional wall-normal coordinate for the high-  {and low-}enthalpy  {temporal} boundary layers. The transformations employed include the van Driest \cite{van1951turbulent} (a), Zhang et al. \cite{zhang2012mach} (b), Trettel-Larsson \cite{trettel2016mean} (c), data-driven \cite{volpiani2020data} (d), total-stress-based \cite{griffin2021velocity} (e), and constant-stress-layer version of the total-stress-based \cite{griffin2021velocity} (f). The color of lines indicates the friction Reynolds number $Re_{\tau}$. Two incompressible DNS mean velocity profiles, i.e., from the channel flow (black dashed line) by Lee and Moser \cite{lee2015direct} at $Re_{\tau}\approx5200$ and the ZPG boundary layer (black dotted line) by Sillero et al. \cite{sillero2013one} at $Re_{\tau}\approx2000$, are also shown for comparisons.} The van Driest \cite{van1951turbulent} transformation fails, which may be due to the heat flux at the wall. There is an obvious increase of the log-law intercept for the van Driest \cite{van1951turbulent} and Zhang et al. \cite{zhang2012mach} transformed velocity profiles, and similar behavior has also been noticed by Duan and Mart{\'i}n \cite{duan2011p4direct}, and Maeder \cite{maeder2000numerical} for the cold-wall boundary layers. The data-driven approach \cite{volpiani2020data} and the total-stress-based transformation perform well in the viscous sublayer, but their performance deteriorates significantly in the logarithmic region. It is worth mentioning that there are two cases, i.e., LowH\_M3 and Wedge35noncata (the readers are referred to \cite{duan2011p4direct} for details), whose transformed profiles from the total-stress-based approach coincide well with the incompressible reference by Sillero et al. \cite{sillero2013one}. Overall speaking, the Trettel-Larsson \cite{trettel2016mean} transformation works the best among all the others.

 {In Fig.~\ref{fig:vel_renzo}, the velocity transformations are shown for the high-enthalpy spatial boundary layer of di Renzo and Urzay \cite{renzo2021direct}. Contrary to the temporal boundary layer, the data-driven and total-stress-based transformations are in better agreement with the incompressible reference than the Trettel-Larsson transformation is. Like the temporal boundary layer, the constant-stress-layer assumption does not visibly affect the results, so $u_{t}^+$ is not shown for brevity. Also, the van Driest transformation is plotted by di Renzo and Urzay \cite{renzo2021direct}. It is worth noting that Passiatore \cite{passiatore2021} performed a direct numerical simulation of a hypersonic, spatially developing, fully turbulent boundary layer with thermochemical effects. They also observed that the total-stress-based transformation out-performs the Trettel-Larsson transformation in the logarithmic region.}

 {Returning to the temporal boundary layer, in Fig.~\ref{fig:diag_duan}, three types of non-dimensional mean shear multiplied by the semi-local wall-normal coordinate are plotted with respect to the semi-local wall-normal coordinate, where the one derived from Trettel-Larsson \cite{trettel2016mean} is in green, the one based on turbulence quasi-equilibrium in cyan (see Eq.~(\ref{eq:T1})), the one from the total-stress-based \cite{griffin2021velocity} transformation in black (see Eq.~(\ref{eq:T4})), and the last one $y^+(\partial u^+/\partial y^+)$ for the incompressible channel flow by Lee and Moser \cite{lee2015direct} at $Re_{\tau}\approx5200$ is in magenta.} The non-dimensional mean shear derived by Trettel-Larsson \cite{trettel2016mean} has a slight downward translation in the viscous sublayer and an upward translation in the buffer layer and the logarithmic region compared with the incompressible reference. And these errors cancel out during the integration (see Eq.~(\ref{eq:TL4})), leading to the success of the Trettel-Larsson \cite{trettel2016mean} transformation in the log region. The total-stress-based transformation is inaccurate in the log region because the modified equilibrium transformation is failing. This transformation makes three assumptions which are assessed individually. (1) In Fig.~\ref{fig:diss_Duan}(a), the quantity $\epsilon^{\ddagger}=\epsilon^+/\mu^+$ does not collapse to the incompressible reference data in the log region, which disrupts the first assumption. (2) In Fig.~\ref{fig:diss_Duan}(b), the ratio of the modified production and the dissipation does not collapse to the incompressible reference data, which invalidates the assumption of the Mach invariance of this ratio. (3) In Fig.~\ref{fig:RS_Duan} (the Reynolds-averaged Reynolds stress is approximated as $\overline{\rho}^+\overline{u'v'}^+$ since only this statistic is available for this database), the Reynolds shear stress is also not Mach invariant in the log region; this is another reason why the modified equilibrium transformation (and thus also the one based on the total shear stress) is failing.  {The disagreement of these three non-dimensional turbulence statistics with incompressible data suggests that the non-dimensionalizations do not accurately account for high-enthalpy effects. Specifically, the near-wall non-dimensional dissipation is much lower than the incompressible data.}
%%%%%%%%%%%%%%%%%%%%%%%%%%%%%
\subsection{Flows at supercritical pressure}
At supercritical pressure, the distinction between the gas and liquid phase blurs, and the flow changes continuously from the liquid phase to the gas phase. The thermodynamic and molecular transport properties have very large gradients near the pseudocritical temperature, known as pseudo-boiling phenomena. These real-fluid effects have inevitable impacts on the near-wall turbulence dynamics and the boundary-layer velocity profile. Additionally, the diabatic wall condition also influences the flow field. As a consequence, it is much more challenging to propose an appropriate near-wall scaling for supercritical flows than for ideal-gas boundary layers.
%%%%%%%%%%%%%
\subsubsection{Channel flows at supercritical pressure}
\begin{figure*}
    \centering
    \includegraphics[width=0.9\linewidth]{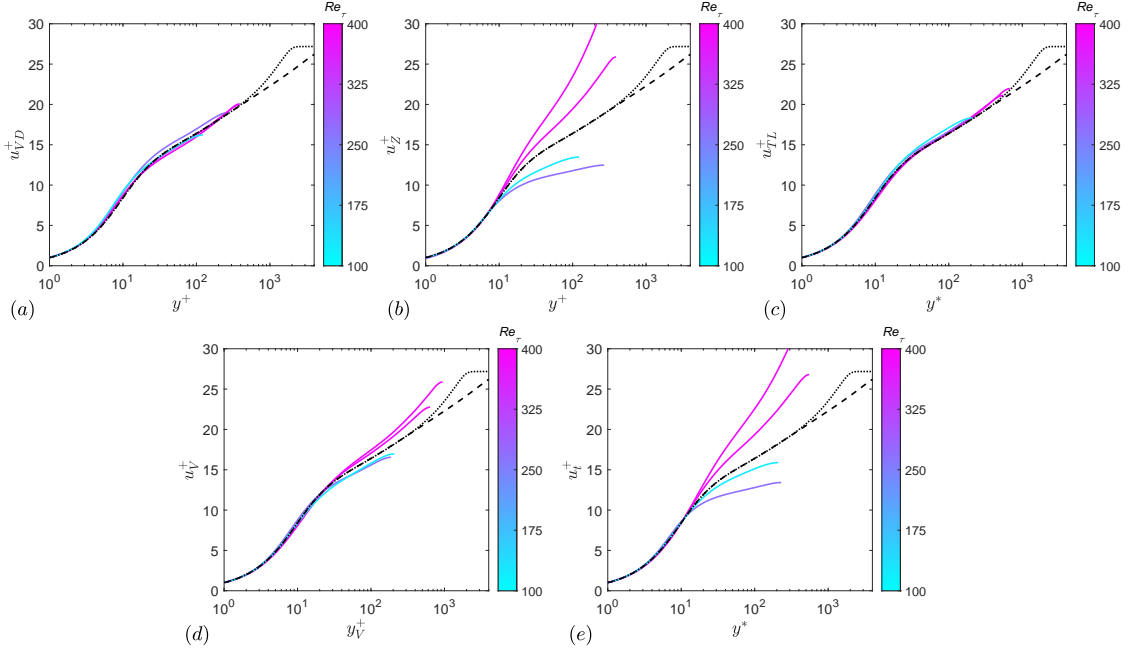}
    \caption{ {Transformed velocity distributions versus the non-dimensional wall-normal coordinate. The database is for channel flows at supercritical pressure by Toki et al. \cite{toki2020velocity}.}}
    %\caption{Transformed velocity distributions versus the non-dimensional wall-normal coordinate with the van Driest \cite{van1951turbulent} (a), Zhang et al. \cite{zhang2012mach} (b), Trettel-Larsson \cite{trettel2016mean} (c), data-driven \cite{volpiani2020data} (d), and total-stress-based \cite{griffin2021velocity} (e) transformation. The database is for channel flows at supercritical pressure by Toki et al. \cite{toki2020velocity}. The color of lines indicates the friction Reynolds number $Re_{\tau}$. Two incompressible DNS mean velocity profiles, i.e., from the channel flow (black dashed line) by Lee and Moser \cite{lee2015direct} at $Re_{\tau}\approx5200$ and the ZPG boundary layer (black dotted line) by Sillero et al. \cite{sillero2013one} at $Re_{\tau}\approx2000$, are also shown for comparisons.}
    \label{fig:vel_toki}
\end{figure*}
\begin{figure*}
    \centering
    \includegraphics[width=0.45\linewidth]{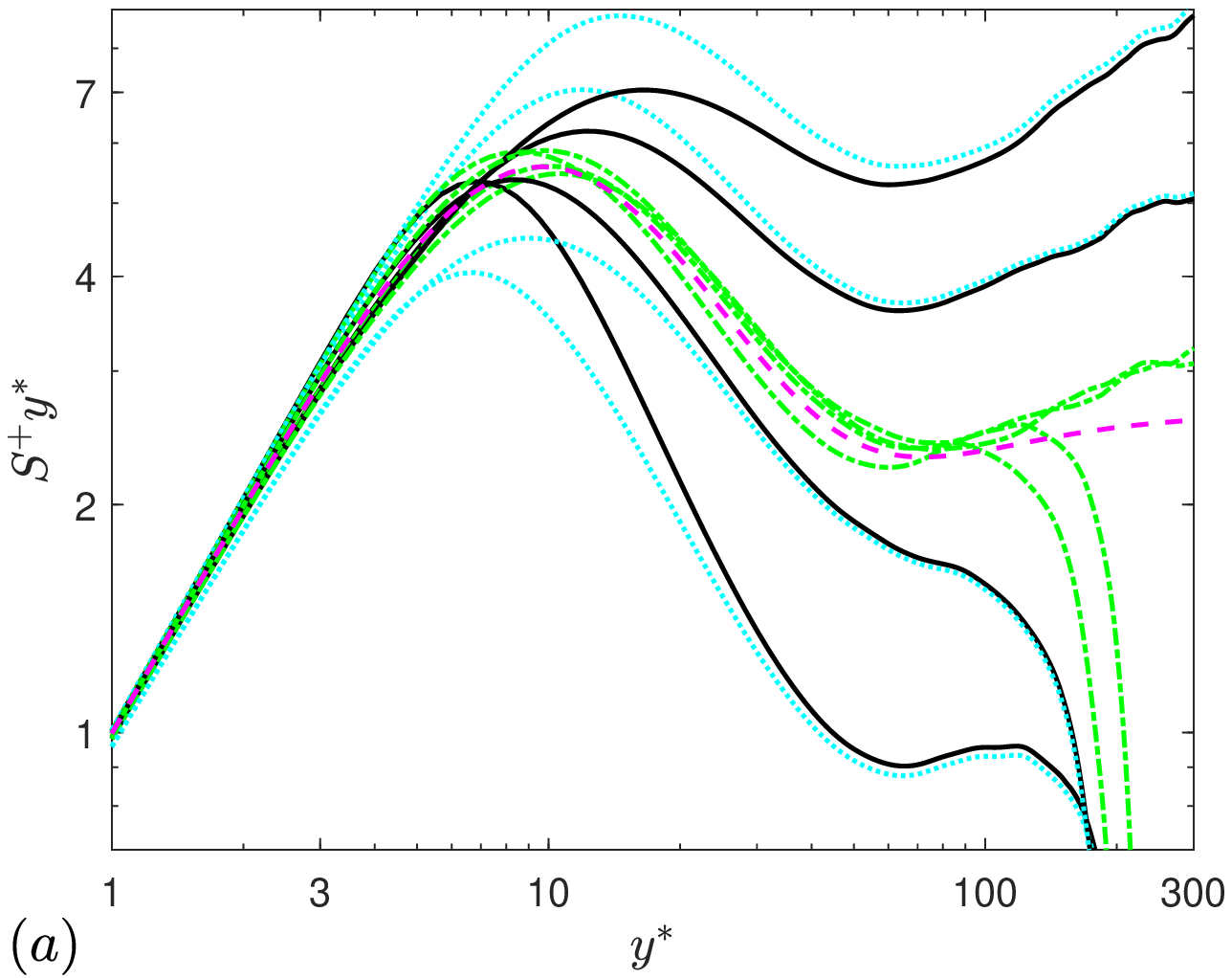}
    \includegraphics[width=0.45\linewidth]{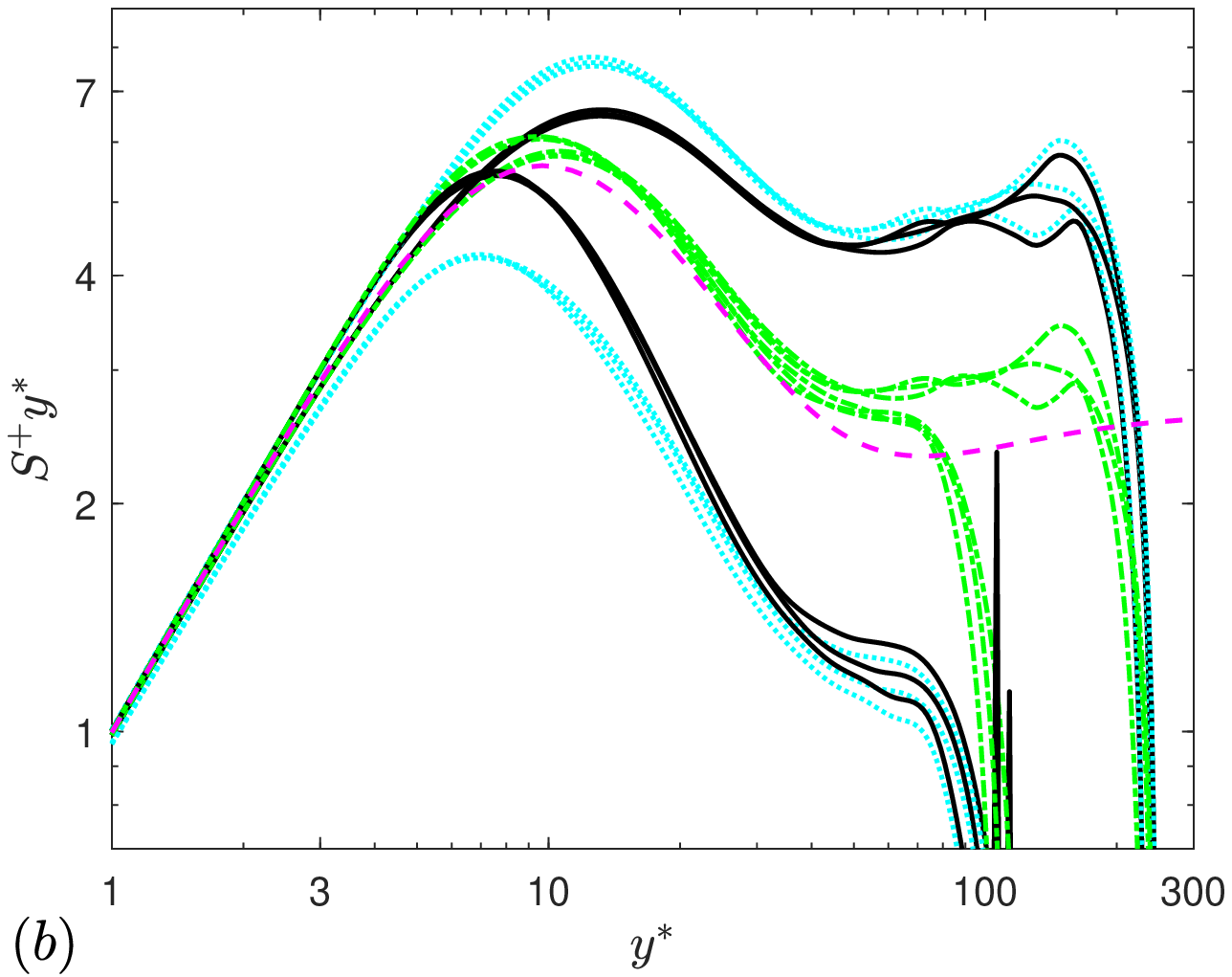}
    \caption{ {Three types of non-dimensional mean shear multiplied by the semi-local wall-normal coordinate are plotted with respect to the semi-local wall-normal coordinate. (a) utilizes the DNS data by Toki et al. \cite{toki2020velocity} and (b) Wan et al. \cite{wan2020mean}.}}
    %\caption{Three types of non-dimensional mean shear multiplied by the semi-local wall-normal coordinate are plotted with respect to the semi-local wall-normal coordinate, where the one derived from Trettel-Larsson \cite{trettel2016mean} is in green, the one based on turbulence quasi-equilibrium in cyan (see Eq.~(\ref{eq:T1})), the one from the total-stress-based \cite{griffin2021velocity} transformation in black (see Eq.~(\ref{eq:T4})), and the last one $y^+(\partial u^+/\partial y^+)$ for the incompressible channel flow by Lee and Moser \cite{lee2015direct} at $Re_{\tau}\approx5200$ in magenta. (a) utilizes the DNS data by Toki et al. \cite{toki2020velocity} and (b) Wan et al. \cite{wan2020mean}.}
    \label{fig:diag_toki_wan}
\end{figure*}
\begin{figure*}
    \centering
    \includegraphics[width=0.45\linewidth]{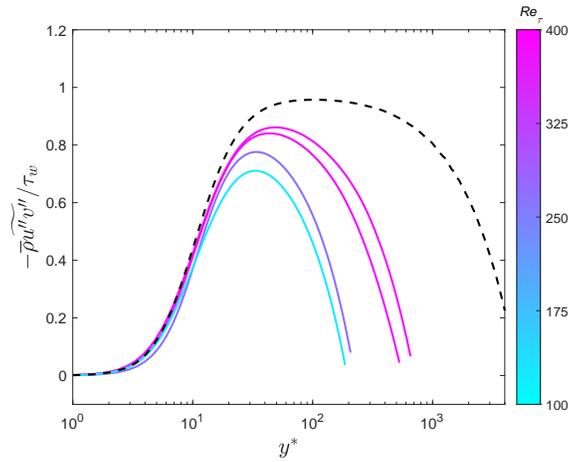}
    \caption{The Reynolds shear stress normalized by the wall stress for the database of channel flows at supercritical pressure by Toki et al. \cite{toki2020velocity}. The incompressible channel flow by Lee and Moser \cite{lee2015direct} at $Re_{\tau}\approx5200$ is employed as reference in the dashed black line.}
    \label{fig:RS_toki}
\end{figure*}

 {In Fig.~\ref{fig:vel_toki}, transformed velocity distributions versus the non-dimensional wall-normal coordinate are shown for the van Driest \cite{van1951turbulent} (a), Zhang et al. \cite{zhang2012mach} (b), Trettel-Larsson \cite{trettel2016mean} (c), data-driven \cite{volpiani2020data} (d), and total-stress-based \cite{griffin2021velocity} (e) transformations. The color of lines indicates the friction Reynolds number $Re_{\tau}$. Two incompressible DNS mean velocity profiles, i.e., from the channel flow (black dashed line) by Lee and Moser \cite{lee2015direct} at $Re_{\tau}\approx5200$ and the ZPG boundary layer (black dotted line) by Sillero et al. \cite{sillero2013one} at $Re_{\tau}\approx2000$, are also shown for comparisons.} The performance of various transformations is compared for the database from Toki et al. \cite{toki2020velocity}. Surprisingly, the Trettel-Larsson \cite{trettel2016mean} and van Driest \cite{van1951turbulent} transformed mean velocity profiles show good agreement with the incompressible law of the wall even for supercritical channel flows, especially when the Trettel-Larsson transformation is deployed. The rest three transformations perform well in the viscous sublayer but fail in the buffer layer and the logarithmic region. The Zhang et al. \cite{zhang2012mach} and total-stress-based \cite{griffin2021velocity} transformations fail in the logarithmic region in these cases, which can be further confirmed by checking the diagnostic function in Fig.~\ref{fig:diag_toki_wan}(a).  {The non-dimensional mean shear derived from Trettel-Larsson \cite{trettel2016mean} is in green, the one based on turbulence quasi-equilibrium in cyan (see Eq.~(\ref{eq:T1})), the one from the total-stress-based \cite{griffin2021velocity} transformation in black (see Eq.~(\ref{eq:T4})), and the last one $y^+(\partial u^+/\partial y^+)$ for the incompressible channel flow by Lee and Moser \cite{lee2015direct} at $Re_{\tau}\approx5200$ in magenta.}
In Fig.~\ref{fig:RS_toki}, the Reynolds shear stress profiles are shown to depart from the incompressible reference, which violates an assumption made by Zhang et al. and Griffin et al., and could explain the failure of these transformations. Since data for the dissipation is not available, the other assumptions made by these transformations can not be interrogated for this data set.

\begin{figure*}
    \centering
    \includegraphics[width=0.9\linewidth]{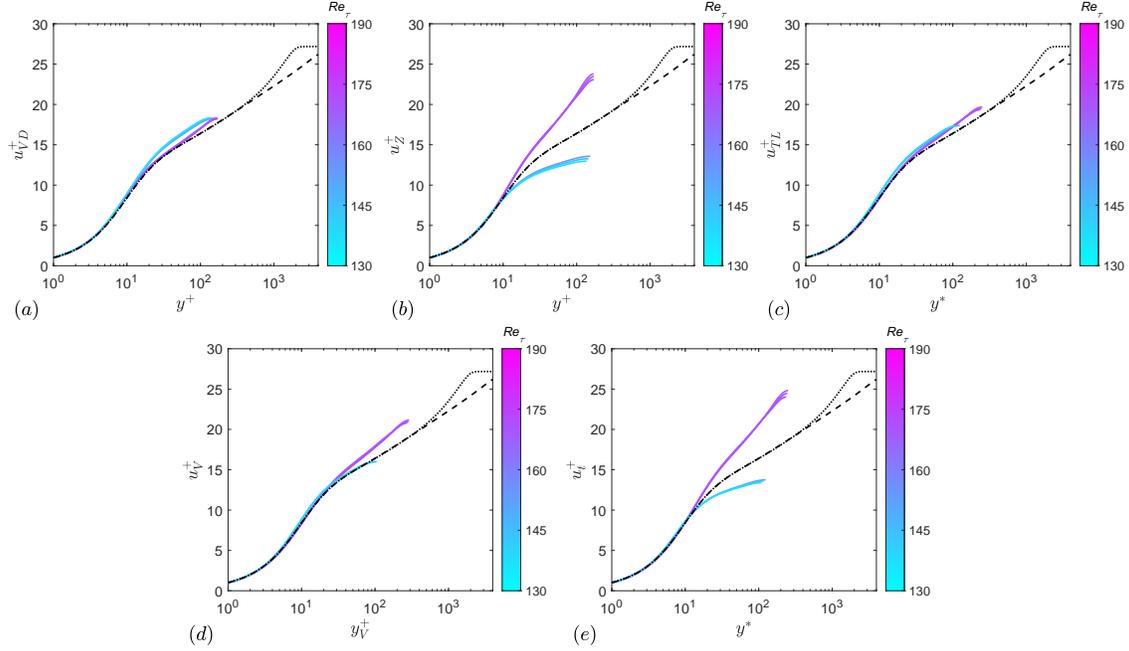}
    \caption{ {Transformed velocity distributions versus the non-dimensional wall-normal coordinate. The database is for channel flows at supercritical pressure by Wan et al. \cite{wan2020mean}.}}
    %\caption{Transformed velocity distributions versus the non-dimensional wall-normal coordinate with the van Driest \cite{van1951turbulent} (a), Zhang et al. \cite{zhang2012mach} (b), Trettel-Larsson \cite{trettel2016mean} (c), data-driven \cite{volpiani2020data} (d), and total-stress-based \cite{griffin2021velocity} (e) transformation. The database is for channel flows at supercritical pressure by Wan et al. \cite{wan2020mean}. The color of lines indicates the friction Reynolds number $Re_{\tau}$. Two incompressible DNS mean velocity profiles, i.e., from the channel flow (black dashed line) by Lee and Moser \cite{lee2015direct} at $Re_{\tau}\approx5200$ and the ZPG boundary layer (black dotted line) by Sillero et al. \cite{sillero2013one} at $Re_{\tau}\approx2000$, are also shown for comparisons.}
    \label{fig:vel_wan}
\end{figure*}

\begin{figure*}
    \centering
    \includegraphics[width=0.45\linewidth]{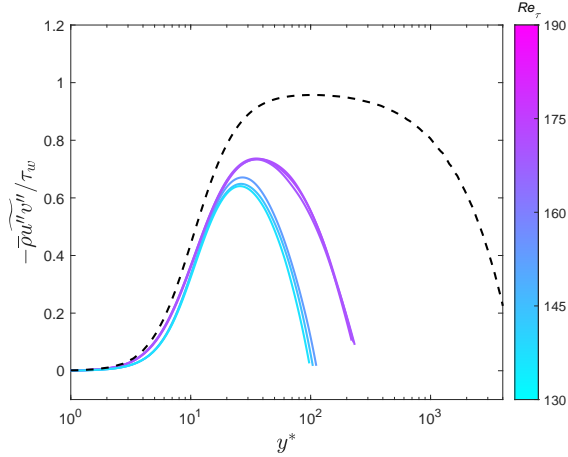}
    \caption{The Reynolds shear stress normalized by the wall stress for the database of channel flows at supercritical pressure by Wan et al. \cite{wan2020mean}. The incompressible channel flow by Lee and Moser \cite{lee2015direct} at $Re_{\tau}\approx5200$ is employed as reference in the dashed black line.}
    \label{fig:RS_wan}
\end{figure*}

 {In Fig.~\ref{fig:vel_wan}, transformed velocity distributions are plotted versus the non-dimensional wall-normal coordinate for the channel flows at supercritical pressure by Wan et al. \cite{wan2020mean}. The velocity transformations plotted include that of van Driest \cite{van1951turbulent} (a), Zhang et al. \cite{zhang2012mach} (b), Trettel-Larsson \cite{trettel2016mean} (c), data-driven \cite{volpiani2020data} (d), and total-stress-based \cite{griffin2021velocity} (e) transformations. The color of lines indicates the friction Reynolds number $Re_{\tau}$. Two incompressible DNS mean velocity profiles, i.e., from the channel flow (black dashed line) by Lee and Moser \cite{lee2015direct} at $Re_{\tau}\approx5200$ and the ZPG boundary layer (black dotted line) by Sillero et al. \cite{sillero2013one} at $Re_{\tau}\approx2000$, are also shown for comparisons.}
It is evident that results from cool walls (low $Re_{\tau}$) and hot walls (high $Re_{\tau}$) follow different patterns, which can be seen in Fig.~\ref{fig:vel_toki} as well. The van Driest \cite{van1951turbulent} and Trettel-Larsson \cite{trettel2016mean} transformations perform better, and an increase of the log-law intercept can be observed for hot walls in these two transformations. The data-driven approach works well for the cases with hot walls, but fails for those with cold walls. Velocity profiles obtained through the Zhang's \cite{zhang2012mach} and total-stress-based \cite{griffin2021velocity} transformations collapse to the incompressible law of the wall only in the viscous sublayer, and the non-dimensional mean shears plotted in Fig.~\ref{fig:diag_toki_wan}(b) imply their failure in the logarithmic region. As shown in Fig.~\ref{fig:RS_wan}, a conclusion similar to that for the database from Toki et al. \cite{toki2020velocity} can be obtained.

%%
%%%%%%%%%%%%%
\subsubsection{Boundary layers at supercritical pressure}
\begin{figure*}
    \centering
    \includegraphics[width=0.9\linewidth]{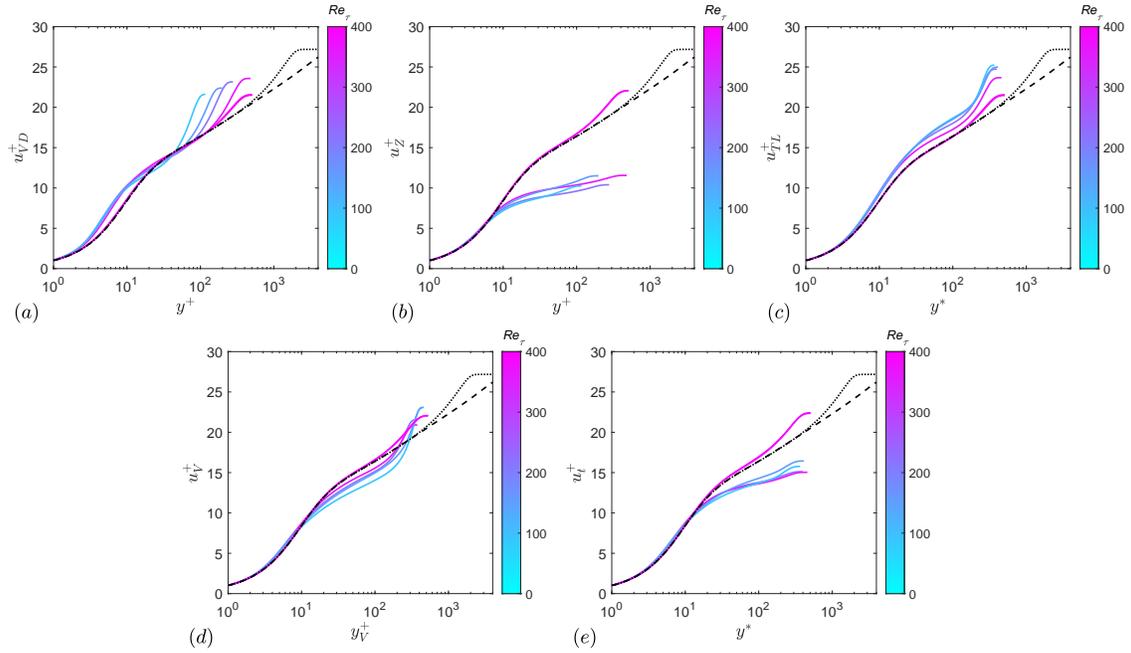}
    \caption{ {Transformed velocity distributions versus the non-dimensional wall-normal coordinate. The database is for boundary layers at supercritical pressure by Kawai \cite{kawai2019heated}.}}
    %\caption{Transformed velocity distributions versus the non-dimensional wall-normal coordinate with the van Driest \cite{van1951turbulent} (a), Zhang et al. \cite{zhang2012mach} (b), Trettel-Larsson \cite{trettel2016mean} (c), data-driven \cite{volpiani2020data} (d), and total-stress-based \cite{griffin2021velocity} (e) transformation. The database is for boundary layers at supercritical pressure by Kawai \cite{kawai2019heated}. The color of lines indicates the friction Reynolds number $Re_{\tau}$. Two incompressible DNS mean velocity profiles, i.e., from the channel flow (black dashed line) by Lee and Moser \cite{lee2015direct} at $Re_{\tau}\approx5200$ and the ZPG boundary layer (black dotted line) by Sillero et al. \cite{sillero2013one} at $Re_{\tau}\approx2000$, are also shown for comparisons.}
    \label{fig:vel_kawai}    
\end{figure*}

\begin{figure*}
    \centering
    \includegraphics[width=0.45\linewidth]{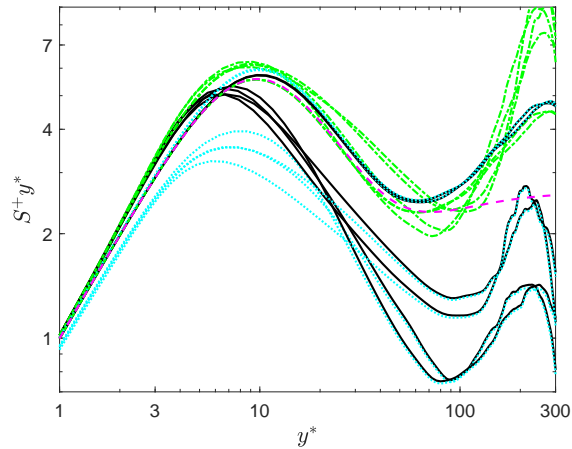}
    \caption{ {Three types of non-dimensional mean shear multiplied by the semi-local wall-normal coordinate are plotted with respect to the semi-local wall-normal coordinate.}}
    %\caption{Three types of non-dimensional mean shear multiplied by the semi-local wall-normal coordinate are plotted with respect to the semi-local wall-normal coordinate, where the one derived from Trettel-Larsson \cite{trettel2016mean} is in green, the one based on turbulence quasi-equilibrium in cyan (see Eq.~(\ref{eq:T1})), the one from the total-stress-based \cite{griffin2021velocity} transformation in black (see Eq.~(\ref{eq:T4})), and the last one $y^+(\partial u^+/\partial y^+)$ for the incompressible channel flow by Lee and Moser \cite{lee2015direct} at $Re_{\tau}\approx5200$ in magenta. The DNS data by Kawai \cite{kawai2019heated} is used.}
    %and (b) Wenzel et al. \cite{wenzel2019self}
    \label{fig:diag_kawai}
\end{figure*}

\begin{figure*}
    \centering
    \includegraphics[width=0.45\linewidth]{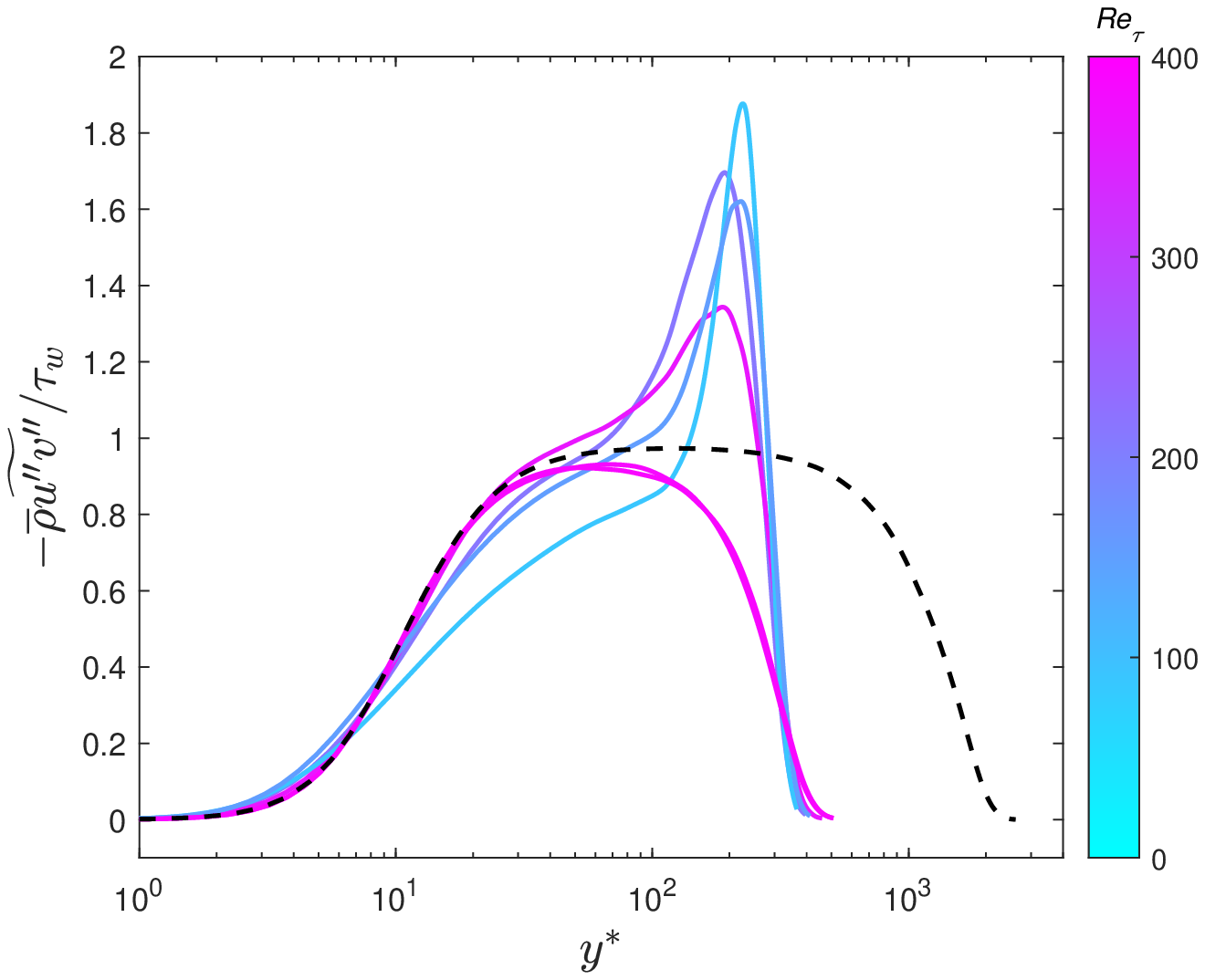}
    \caption{The Reynolds shear stress normalized by the wall stress for the database of turbulent boundary layers at supercritical pressure by Kawai \cite{kawai2019heated}. The incompressible boundary layer by Sillero et al. \cite{sillero2013one} at $Re_{\tau}\approx2000$ is employed as reference in the dashed black line.}
    \label{fig:RS_kawai}
\end{figure*}

 {In Fig.~\ref{fig:vel_kawai}, the transformed velocity distributions are plotted versus the non-dimensional wall-normal coordinate for boundary layer flows at supercritical pressure. The velocity transformations include the van Driest \cite{van1951turbulent} (a), Zhang et al. \cite{zhang2012mach} (b), Trettel-Larsson \cite{trettel2016mean} (c), data-driven \cite{volpiani2020data} (d), and total-stress-based \cite{griffin2021velocity} (e) transformations.
The color of lines indicates the friction Reynolds number $Re_{\tau}$. Two incompressible DNS mean velocity profiles, i.e., from the channel flow (black dashed line) by Lee and Moser \cite{lee2015direct} at $Re_{\tau}\approx5200$ and the ZPG boundary layer (black dotted line) by Sillero et al. \cite{sillero2013one} at $Re_{\tau}\approx2000$, are also shown for comparisons. There are two cases which can be mapped to the incompressible law of the wall by all the transformations tested.}
The wall conditions in these two cases are unheated such that the flow temperature remains below the pseudo-critical temperature, and the pseudo-boiling phenomenon does not appear. For the heated cases, the Zhang et al. \cite{zhang2012mach}, data-driven \cite{volpiani2020data}, and total-stress-based transformations only work in the viscous sublayer, and the performance of all the transformations tested deteriorates in the logarithmic region. 
 {In Fig.~\ref{fig:diag_kawai}, three types of non-dimensional mean shear multiplied by the semi-local wall-normal coordinate are plotted with respect to the semi-local wall-normal coordinate. The non-dimensionalization implied by the Trettel-Larsson \cite{trettel2016mean} transformation is in green, the one based on turbulence quasi-equilibrium in cyan (see Eq.~(\ref{eq:T1})), the one from the total-stress-based \cite{griffin2021velocity} transformation in black (see Eq.~(\ref{eq:T4})), and the last one $y^+(\partial u^+/\partial y^+)$ for the incompressible channel flow by Lee and Moser \cite{lee2015direct} at $Re_{\tau}\approx5200$ in magenta. The DNS data by Kawai \cite{kawai2019heated} is used.
}
Fig.~\ref{fig:diag_kawai} also shows that none of the proposed non-dimensional mean shears collapses to the incompressible reference in the entire inner region. The failure of the van Driest \cite{van1951turbulent}, Trettel-Larsson \cite{trettel2016mean}, and Zhang et al. \cite{zhang2012mach} transformation may be due to the effect of heat transfer at the wall since these transformations also do not perform well in  {perfect} gas boundary layers with heat transfer \cite{griffin2021velocity}. Also, the pseudo-boiling phenomenon could be a source of error, and it might compound with the aforementioned error. The data-driven approach \cite{volpiani2020data} and the total-stress-based transformation \cite{griffin2021velocity} fail because of the influence of the pseudo-boiling phenomenon, since they show excellent performance in canonical diabatic boundary layers without supercritical effects. In Fig.~\ref{fig:RS_kawai}, the Reynolds shear stress profiles are shown to depart from the incompressible reference, which violates an assumption made by Zhang et al. and Griffin et al., and could explain the failure of these transformations.
 {Unlike canonical boundary layers, the Favre-averaged Reynolds-shear stress data exceeds unity for some cases due to the excessive positive contribution from the mean-density-related term. Large changes in density arise in cases where wall heating leads to transcritical thermal conditions (see page 585 in \cite{kawai2019heated} for details).  As a result, the Reynolds-shear stress exceeds unity in only the cases with wall heating.}


In summary, although the Trettel-Larsson \cite{trettel2016mean} transformation is widely recognized as one of the most appropriate transformations for channel flows, it is unexpected that it is also successful in supercritical channel flows. For boundary layers at supercritical pressure, none of the transformations is able to collapse these mean velocity profiles and the appropriate transformation still remains unclear.

%%%%%%%%%%%%%%%%%%%%%%%%%%%
\subsection{Boundary layers with pressure gradients}
\begin{figure*}
    \centering
    \includegraphics[width=0.9\linewidth]{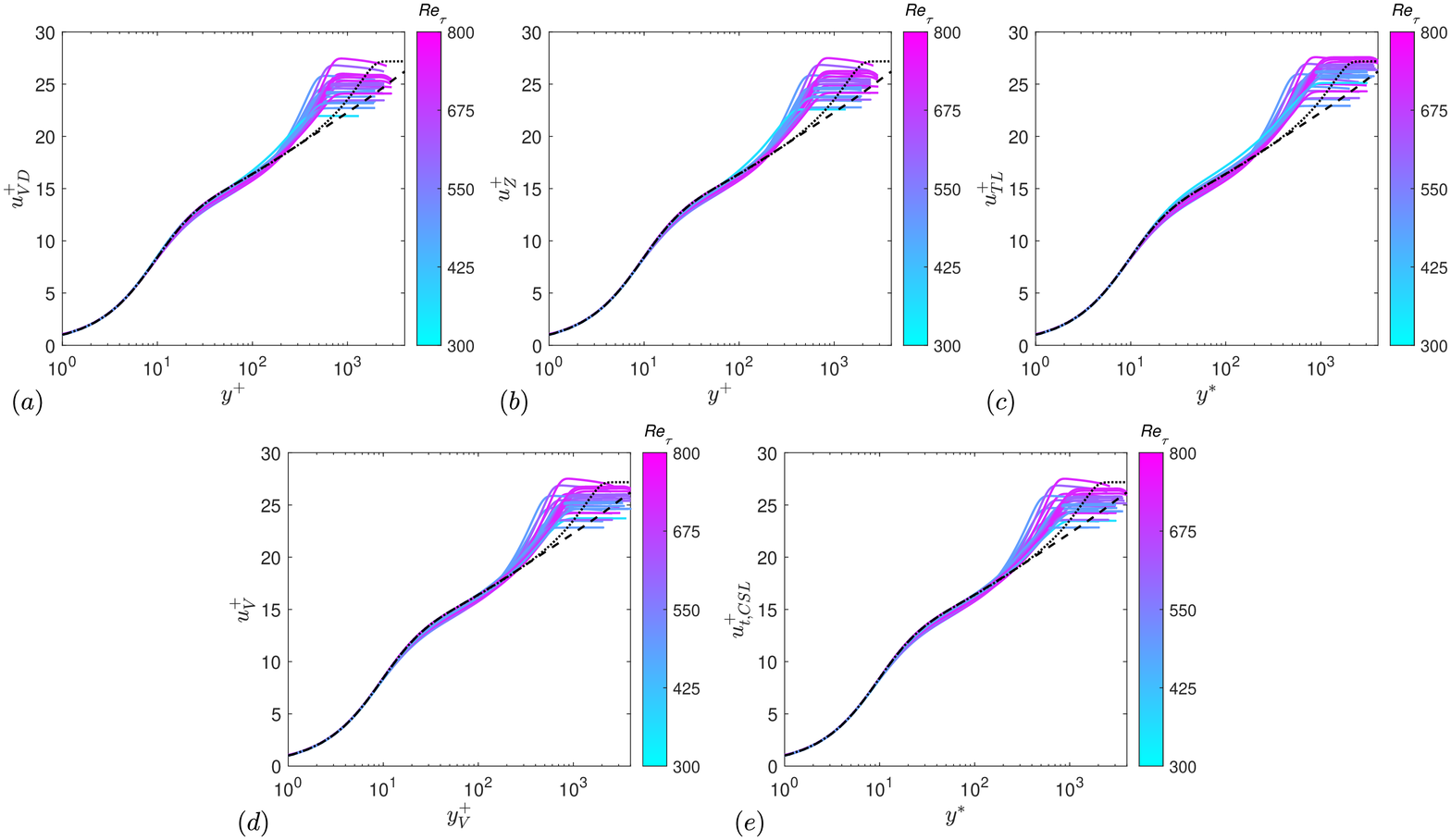}
    \caption{ {Transformed velocity distributions versus the non-dimensional wall-normal coordinate. The database is for boundary layers with the pressure gradient by Wenzel et al. \cite{wenzel2019self}.}}
    %\caption{Transformed velocity distributions versus the non-dimensional wall-normal coordinate with the van Driest \cite{van1951turbulent} (a), Zhang et al. \cite{zhang2012mach} (b), Trettel-Larsson \cite{trettel2016mean} (c), data-driven \cite{volpiani2020data} (d), and total-stress-based \cite{griffin2021velocity} (e) transformation. The database is for boundary layers with the pressure gradient by Wenzel et al. \cite{wenzel2019self}. The color of lines indicates the friction Reynolds number $Re_{\tau}$. Two incompressible DNS mean velocity profiles, i.e., from the channel flow (black dashed line) by Lee and Moser \cite{lee2015direct} at $Re_{\tau}\approx5200$ and the ZPG boundary layer (black dotted line) by Sillero et al. \cite{sillero2013one} at $Re_{\tau}\approx2000$, are also shown for comparisons.}
    \label{fig:vel_Wenzel}
\end{figure*}
\begin{figure*}
    \centering
    \includegraphics[width=0.45\linewidth]{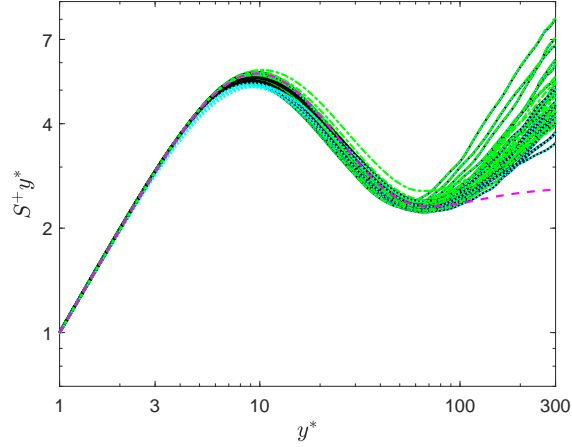}
    \caption{ {Three types of non-dimensional mean shear multiplied by the semi-local wall-normal coordinate are plotted with respect to the semi-local wall-normal coordinate. The DNS data of turbulent boundary layers by Wenzel et al. \cite{wenzel2019self} is used.}}
    %\caption{Three types of non-dimensional mean shear multiplied by the semi-local wall-normal coordinate are plotted with respect to the semi-local wall-normal coordinate, where the one derived from Trettel-Larsson \cite{trettel2016mean} is in green, the one based on turbulence quasi-equilibrium in cyan (see Eq.~(\ref{eq:T1})), the one from the total-stress-based \cite{griffin2021velocity} transformation in black (see Eq.~(\ref{eq:T4})), and the last one $y^+(\partial u^+/\partial y^+)$ for the incompressible channel flow by Lee and Moser \cite{lee2015direct} at $Re_{\tau}\approx5200$ in magenta. The DNS data of turbulent boundary layers by Wenzel et al. \cite{wenzel2019self} is used.}
    \label{fig:diag_Wenzel}
\end{figure*}
 {In Fig.~\ref{fig:vel_Wenzel}, transformed velocity distributions are plotted versus the non-dimensional wall-normal coordinate for the database is for boundary layers with the pressure gradient by Wenzel et al. \cite{wenzel2019self}. The transformaitons include the van Driest \cite{van1951turbulent} (a), Zhang et al. \cite{zhang2012mach} (b), Trettel-Larsson \cite{trettel2016mean} (c), data-driven \cite{volpiani2020data} (d), and total-stress-based \cite{griffin2021velocity} (e) transformations. The color of lines indicates the friction Reynolds number $Re_{\tau}$. Two incompressible DNS mean velocity profiles, i.e., from the channel flow (black dashed line) by Lee and Moser \cite{lee2015direct} at $Re_{\tau}\approx5200$ and the ZPG boundary layer (black dotted line) by Sillero et al. \cite{sillero2013one} at $Re_{\tau}\approx2000$, are also shown for comparisons.}
The performance of all the considered transformations is similar for boundary layers with pressure gradients, and good agreement is observed between the transformed velocity distributions and the incompressible law of the wall, although the log-law intercept varies in a small range due to the pressure gradient effects characterized by the Rotta-Clauser parameter $\beta_k$. Such a good collapse can be further confirmed by Fig.~\ref{fig:diag_Wenzel}, which shows that the  {three} non-dimensional mean shears agree well with the incompressible reference. 
 {The three non-dimensional mean shears include the one derived from the Trettel-Larsson \cite{trettel2016mean} transformation in green, the one based on turbulence quasi-equilibrium in cyan (see Eq.~(\ref{eq:T1})), the one from the total-stress-based \cite{griffin2021velocity} transformation in black (see Eq.~(\ref{eq:T4})), and the last one $y^+(\partial u^+/\partial y^+)$ for the incompressible channel flow by Lee and Moser \cite{lee2015direct} at $Re_{\tau}\approx5200$ in magenta.}
It is worth mentioning that the pressure gradients are relatively weak for these cases and, moreover, the wall boundary condition is adiabatic, which further simplifies the near-wall dynamics \cite{wenzel2019self}. Such an observation indicates that, for wall-bounded turbulence with a weak pressure gradient, the transformations developed for ZPG boundary layers can be sufficiently accurate.
%
%%%%%%%%%%%%% TABLE1 %%%%%%%%%%%%%%
\begin{table}[!htbp]
\caption{ {Summary of the performance in the viscous sublayer of various velocity transformations.}\label{tab:1}}
\centering
\begin{tabular}{cccccc}
\hline
\hline
Transformations & BLHE  {\cite{duan2011p4direct,renzo2021direct}} & CSPT \cite{toki2020velocity} & CSPW \cite{wan2020mean} & BLSP \cite{kawai2019heated} & BLPG \cite{wenzel2019self,gibis2019self}\\ 
\hline
\hline
van Driest \cite{van1951turbulent} & $\times$  & $\checkmark$ & $\checkmark$ & $\times$ & $ \checkmark$\\
\hline
Zhang et al. \cite{zhang2012mach} & $\times$  & $\checkmark$ & $\checkmark$ & $\checkmark$ & $\checkmark$\\
\hline
Trettel and Larsson \cite{trettel2016mean} & $\checkmark$ {\color{black}, $\checkmark$} & $\checkmark$ & $\checkmark$ &  $\times$ & $\checkmark$ \\
\hline
Volpiani et al. \cite{volpiani2020data} & $\checkmark$ {\color{black}, $\checkmark$} & $\checkmark$ & $\checkmark$ & $\checkmark$ & $\checkmark$ \\
\hline
Griffin et al. \cite{griffin2021velocity} & $\checkmark$ {\color{black}, $\checkmark$} & $\checkmark$ & $\checkmark$ & $\checkmark$ & $\checkmark$ \\
\hline
\end{tabular}\\
\footnotesize{ {$\checkmark$ indicates adequate performance and $\times$ denotes unsatisfactory performance. BLHE denotes high-enthalpy  {and low-enthalpy} turbulent  {temporal and spatial} boundary layers, CSPT denotes channel flows at supercritical pressure, CSPW denotes channel flows at supercritical pressure, BLSP denotes boundary layers at supercritical pressure, and BLPG denotes boundary layers with pressure gradients}}\\

%\caption{Summary of the performance in the viscous sublayer of various velocity transformations. $\checkmark$ indicates adequate performance and $\times$ denotes unsatisfactory performance. BLHE denotes high-enthalpy turbulent boundary layers from Duan and Mart{\'i}n \cite{duan2011p4direct}, CSPT denotes channel flows at supercritical pressure from Toki et al. \cite{toki2020velocity}, CSPW denotes channel flows at supercritical pressure from Wan et al. \cite{wan2020mean}, BLSP denotes boundary layers at supercritical pressure from Kawai \cite{kawai2019heated}, and BLPG denotes boundary layers with pressure gradients from Wenzel et al. \cite{wenzel2019self} and Gibis et al. \cite{gibis2019self}.\label{tab:1}}
\end{table}
%%%%%%%%%%%%% END TABLE1 %%%%%%%%%%%%%%
%
%%%%%%%%%%%%% TABLE2 %%%%%%%%%%%%%%
\begin{table}[!htbp]
\centering
\caption{ {Summary of the performance in the logarithmic region of various velocity transformations.}\label{tab:2}}
\begin{tabular}{cccccc}
\hline
\hline
Transformations  & BLHE  {\cite{duan2011p4direct,renzo2021direct}} & CSPT \cite{toki2020velocity} & CSPW \cite{wan2020mean} & BLSP \cite{kawai2019heated} & BLPG \cite{wenzel2019self,gibis2019self}\\ 
\hline
\hline
van Driest \cite{van1951turbulent} & $\times$ & $\checkmark$ & $\checkmark$ & $\times$ & $\checkmark$ \\
\hline
Zhang et al. \cite{zhang2012mach} & $\times$ & $\times$ & $\times$ & $\times$ & $\checkmark$ \\
\hline
Trettel and Larsson \cite{trettel2016mean} & $\checkmark$ {\color{black},} $\times$ & $\checkmark$ & $\checkmark$ & $\times$ & $\checkmark$ \\
\hline
Volpiani et al. \cite{volpiani2020data} & $\times$ {\color{black}, $\checkmark$} & $\times$ & $\times$ & $\times$ & $\checkmark$ \\
\hline
Griffin et al. \cite{griffin2021velocity} & $\times$ {\color{black}, $\checkmark$} & $\times$ & $\times$ & $\times$ & $\checkmark$ \\
\hline
\end{tabular}\\
\footnotesize{ {$\checkmark$ indicates adequate performance and $\times$ denotes unsatisfactory performance. BLHE denotes high-enthalpy  {and low-enthalpy} turbulent  {temporal and spatial} boundary layers, CSPT denotes channel flows at supercritical pressure, CSPW denotes channel flows at supercritical pressure, BLSP denotes boundary layers at supercritical pressure, and BLPG denotes boundary layers with pressure gradients.}}
%\caption{Summary of the performance in the  logarithmic region of various velocity transformations. $\checkmark$ indicates adequate performance and $\times$ denotes unsatisfactory performance. BLHE denotes high-enthalpy turbulent boundary layers from Duan and Mart{\'i}n \cite{duan2011p4direct}, CSPT denotes channel flows at supercritical pressure from Toki et al. \cite{toki2020velocity}, CSPW denotes channel flows at supercritical pressure from Wan et al. \cite{wan2020mean}, BLSP denotes boundary layers at supercritical pressure from Kawai \cite{kawai2019heated}, and BLPG denotes boundary layers with pressure gradients from Wenzel et al. \cite{wenzel2019self} and Gibis et al. \cite{gibis2019self}.\label{tab:2}}
\end{table}
%%%%%%%%%%%%% END TABLE2 %%%%%%%%%%%%%%
%
%%%%%%%%%%%%%%%%%%%%%%%%%%%%%%%%%%%%%%%%%%%%%%%%%%%%%%%%%%%%%%
\section{Conclusions}\label{conc}
The objective of this paper is to assess several popular compressible velocity transformations in non-canonical flows including boundary layers with high-enthalpy thermochemical effects wall-bounded turbulent flows at supercritical pressures, and boundary layers with streamwise pressure gradients. Specifically, the following transformations are analyzed, i.e., the van Driest transformation \cite{van1951turbulent}, the transformation of Zhang et al. \cite{zhang2012mach}, the Trettel-Larsson transformation \cite{trettel2016mean}, the data-driven transformation \cite{volpiani2020data}, and the total-stress-based transformation \cite{griffin2021velocity}. The performance of these methods has been summarized in Tables~\ref{tab:1} and~\ref{tab:2}.

The Trettel-Larsson \cite{trettel2016mean} transformation works well for the high-  {and low-}enthalpy  {temporal} boundary-layer flows by Duan and Mart{\'i}n \cite{duan2011p4direct} because the discrepancy between its non-dimensional mean shear and the incompressible reference cancels out during integration. On the other hand, the data-driven approach \cite{volpiani2020data} and the total-stress-based transformation \cite{griffin2021velocity} are approximately valid only in the viscous sublayer.  {For the high-enthalpy spatial boundary layer of di Renzo and Urzay \cite{renzo2021direct}, the conclusion is different. The total-stress-based and data-driven transformations collapse the data well even in the log region and the Trettel-Larsson transformation is only accurate in the viscous sublayer. However, more data is required to further substantiate these observations; as such, they should be regarded as preliminary.}

For supercritical flows, the significant gradient of fluid properties near the pseudocritical temperature results in real-fluid effects. And the interaction of real-fluid effects with the wall-bounded turbulence dynamics remains unclear, which increases the difficulty of proposing an appropriate transformation. 

The Trettel-Larsson \cite{trettel2016mean} transformation works the best in supercritical channel flows and followed by the van Driest \cite{van1951turbulent} transformation. The performance deterioration of the total-stress-based \cite{griffin2021velocity} and Zhang et al. \cite{zhang2012mach} transformation in the log layer may be due to the failure of the assumption of quasi-equilibrium, the Mach-number-invariance of the dissipation profile, or the Mach-number-invariance of the Reynolds shear stress. %%Additional data is required to distinguish these potential sources of error.

For supercritical boundary layers, none of the transformations mentioned is valid. Both the heat transfer at the wall and the real-fluid effect could result in the performance deterioration.

All transformations satisfactorily collapse the velocity profiles of the boundary layers with pressure gradients to the incompressible reference. 
This might be due to the relatively weak pressure gradient imposed in the DNS by Wenzel et al. \cite{wenzel2019self} and the adiabatic wall boundary condition. Further validation of flows with stronger pressure gradients and diabatic wall boundary condition is needed.

In summary, all these considered velocity transformations fail to deliver a uniform performance for non-canonical compressible wall-bounded flows in the log region, and a more sophisticated version, which accounts for these different physics, is needed. 
The data-driven transformation \cite{volpiani2020data} and the total-stress-based transformation \cite{griffin2021velocity} function well in the viscous sublayer for all the considered flows.
Nevertheless, the present assessment provides a useful guideline on the deployment of these transformations on various non-canonical flows.

 {These observations might inspire future developments of mean velocity transformations. Since none of the transformations are valid in the buffer and logarithmic layers of supercritical boundary layers with heat transfer, this regime is an opportunity for further research. Each of the existing methods makes assumptions which are apparently not appropriate for this flow. Such assumptions include the constant stress layer, the validity of Prandtl's incompressible mixing length, the balance of production and dissipation, etc.  It may be possible to modify or generalize these physical assumptions by appealing to DNS data. Then, the mean velocity transformations could be reformulated accordingly.}

 {An alternative avenue for future study is to develop data-driven transformations for non-canonical flows. Volpiani et al. \cite{volpiani2020data} proposed a successful mean velocity transformation for subcritical boundary layers with heat transfer and constant specific heats. In principle, data-driven methods could also be applied to the non-canonical cases considered in this work. Since there are only a limited number of high-fidelity databases on non-canonical flows, it is expected that a data-driven method would have only limited generality. That is, it might perform poorly in cases with parameters that are different from existing data. This may motivate additional high-fidelity simulations of non-canonical flows.}

\section*{Acknowledgments}

K.P.G. acknowledges support from the National Defense Science and Engineering Graduate Fellowship and the Stanford Graduate Fellowship. 
L.F. acknowledges the fund from Shenzhen Municipal Central Government Guides Local Science and Technology Development Special Funds Funded Projects (NO. 2021Szvup138).

%%\bibliography{ref.bib}

\end{document}